\documentclass[11pt,a4paper,prd,noshowpacs,noshowkeys,notitlepage,nofootinbib]%
{revtex4}
\pdfoutput=1 
\usepackage[latin1]{inputenc}
\usepackage{ae,aecompl}
\usepackage{amsmath,amssymb,mathrsfs}
\usepackage{graphicx}
\usepackage[usenames,dvipsnames]{color} 
\usepackage{slashed}
\usepackage{hyperref}
\hypersetup{
    colorlinks=true,      	
    linkcolor=Blue,        	
    citecolor=Plum,        	
    filecolor=magenta,      	
    urlcolor=YellowOrange     	
}

\providecommand{\ZZ}{\mathbb{Z}}
\providecommand{\eq}[1]{\begin{equation} #1 \end{equation}}
\providecommand{\mtrx}[1]{\begin{pmatrix} #1 \end{pmatrix}}
\providecommand{\bs}[1]{\boldsymbol{#1}}
\providecommand{\mss}[1]{\mbox{\scriptsize $#1$}}
\providecommand{\tp}{{\mss{\mathsf{T}}}}
\DeclareMathOperator{\diag}{\mathrm{diag}} 
\providecommand{\RR}{\mathbb{R}}
\providecommand{\CC}{\mathbb{C}}
\providecommand{\eqali}[1]{\begin{equation}\begin{aligned} #1
    \end{aligned}\end{equation}}
\providecommand{\hs}[1]{\hspace{#1}}
\providecommand{\aver}[1]{\langle #1 \rangle}
\providecommand{\xlink}[1]
  {\href{http://arxiv.org/abs/#1}{arXiv:#1}}
\providecommand{\bss}{\mathbf{s}}
\providecommand{\tbss}{\tilde{\mathbf{s}}}
\providecommand{\bu}{\mathbf{u}}
\providecommand{\bw}{\mathbf{w}}
\providecommand{\bp}{\mathbf{p}}
\providecommand{\bq}{\mathbf{q}}
\providecommand{\br}{\mathbf{r}}
\providecommand{\tD}{\tilde{D}}
\providecommand{\tphi}{\tilde{\phi}}
\DeclareMathOperator{\rank}{rank}

\DeclareMathOperator{\snf}{SNF}

\providecommand{\tG}{\tilde{G}}
\providecommand{\bd}{\mathbf{d}}

\providecommand{\lag}{\mathscr{L}}

\providecommand{\boundY}{|G_q|_{\rm bound}}
\providecommand{\boundV}{|G_\phi|_{\rm max}}
\providecommand{\full}{\mathrm{full}}

\begin{document}
\title{
Compatible abelian symmetries in N-Higgs-Doublet Models
}
\author{C.~C.~Nishi}
\email{celso.nishi@ufabc.edu.br}
\affiliation{
Universidade Federal do ABC - UFABC, 
09.210-170, Santo André, SP, Brazil
}
\affiliation{Maryland Center for Fundamental Physics, 
University of Maryland, College Park, MD 20742, USA}


\begin{abstract}
We analyze the compatibility between abelian symmetries acting in two different 
sectors of a theory using the Smith Normal Form method.
We focus on N-Higgs-doublet models (NHDMs) and on the compatibility between 
symmetries in the Higgs potential and in the Yukawa interactions, which were 
separately analyzed in previous works.
It is shown that two equal (isomorphic) symmetry groups that act in two separate 
sectors are not necessarily compatible in the whole theory and an upper 
bound is found for the size of the group that can be implemented in the entire NHDM.
We also develop useful techniques to analyze compatibility and extend a symmetry 
from one sector to another.
Consequences to the supersymmetric case are briefly discussed.
\end{abstract}
\maketitle
\section{Introduction}
\label{sec:intro}

Symmetry has always played a crucial role in our understanding of fundamental 
physics. The construction of the current framework -- the Standard Model (SM) of 
particle physics -- has culminated in 2012 with the discovery of the Higgs 
boson\,\cite{higgs.12}, the particle that results from the breaking of the 
electroweak symmetry in its simplest form.
Hence, it was also a successful attempt to probe a hidden (broken) symmetry in 
nature and its breaking mechanism.
However, as we probe higher and higher energies, new symmetries may emerge as key 
ingredients to understand the physics beyond the SM.

As we try to guess which new symmetry governs the physics above the electroweak 
scale, we are also confronted with the question of what is the breaking scale and 
what could be the signatures after breaking.
One old but fruitful example where the symmetry should (usually) be broken at very 
high energies is $B-L$ symmetry, a symmetry that might be linked to the 
smallness of neutrino masses (see, e.g., Ref.\,\cite{b-l} and references 
therein).

In parallel to continuous symmetries, discrete symmetries are also possible 
ingredients with which we can understand flavor (for a review, see e.g. 
Refs.\cite{flavor.groups}) and the stability of dark matter 
(with, e.g., R-parity\,\cite{r-parity} or matter parity\,\cite{matter-parity}).
In the effort to classify and discover useful abelian discrete symmetries, the 
Smith Normal Form (SNF) method has been used successfully in various contexts to
find discrete symmetries arising from the breaking of continuous gauge 
symmetries\,\cite{schieren}, 
find useful R-symmetries in supersymmetric extensions of the SM\,\cite{ratz.14}, 
justify two-zero textures in the neutrino mass matrix 
with symmetries\,\cite{serodio:nu.texture} and classify abelian symmetries 
in multi-Higgs-doublets models\,\cite{abelian:yuk}.
The latter class of models will be the focus of this work.

The N-Higgs-doublet models (NHDMs) are among the most conservative extensions of 
the SM and they can present additional features that are absent in the 
single-Higgs-doublet SM such as spontaneous CP violation\,\cite{lee,weinberg:scpv} 
or geometric CP violation\,\cite{geo.cp}.
In many ways, these new phenomena are possible because the scalar potential has 
more structure to allow different symmetry breaking paths.
Along with more structure comes the possibility of accommodating larger 
symmetries, specially discrete symmetries.
One can for example impose a $\ZZ_2$ symmetry to naturally suppress dangerous 
flavor changing currents for quarks\,\cite{nfc} or obtain a dark matter candidate 
with radiative neutrino mass generation\,\cite{ma}.
The list of all possible symmetries that can be accommodated in the 2HDM is 
short and the groups in it are small\,\cite{2hdm:sym,joao:abelian} 
(more symmetries arise if we allow for accidental symmetries\,\cite{custodial}).
More and larger discrete symmetries are possible in the 3HDM 
potential\,\cite{3hdm:finite}, and all possible breaking patterns were summarized 
recently in Ref.\,\cite{3hdm:break}.
In general, we can accommodate larger symmetries as we add more fields.
If we are restricted to abelian symmetries, the maximal order of the group that can 
be separately implemented in the Higgs potential and in the Yukawa interactions 
were presented in Refs.\,\cite{abelian:V} and \cite{abelian:yuk}, 
respectively.

In this work, we want to extend the methods of \cite{abelian:yuk} to consider the 
\textit{compatibility issues}, i.e., (i) how to analyze if a symmetry acting 
in two sectors of a theory are compatible and (ii) how to extend a given symmetry 
from one sector to another.
As an example of a situation where we want an answer to (i), we know $\ZZ_8$ is 
the largest symmetry we can implement separately in the Higgs potential and in the 
Yukawa interactions for four Higgs doublets\,\cite{abelian:V,abelian:yuk} but it 
is not clear if the \textit{same} symmetry can be valid for the entire theory.
In contrast, the necessity for (ii) arises frequently when a symmetry is 
interesting in one sector of the theory and we want to build the entire theory.
To answer these questions, we need to formulate compatibility conditions in terms 
of the Smith Normal form and develop techniques to check and extend symmetries.

The outline of this work is as follows: in Sec.\,\ref{sec:review} we review the 
method of the Smith Normal Form to analyze discrete and continuous abelian 
symmetries. We formulate the compatibility conditions in the context of 
N-Higgs-doublet models in Sec.\,\ref{sec:match}, although they can be easily 
adapted to any theory with two sectors or more.
Section \ref{sec:max:3h} shows that the groups with maximal order in the potential 
can be extended compatibly to the whole theory for $N=2,3$.
On the other hand, we show in Sec.\,\ref{sec:discard} that the groups of maximal 
order in the Yukawa sector, for $N\ge 4$, cannot be extended to the potential 
without leading to accidental symmetries.
Methods to extend a symmetry from one sector to another are developed in 
Sec.\,\ref{sec:extension} and we conclude in Sec.\,\ref{sec:conclusion}, 
where we also briefly discuss the supersymmetric case.
Auxiliary material can be found in the appendices.

\section{Review of the method}
\label{sec:review}

We review here the method to apply the Smith normal form (SNF) to \textit{analyze} 
the \textit{rephasing symmetries} of a Lagrangian composed of a polynomial of 
fields. More details can be found in Ref.\,\cite{abelian:yuk}.
We just recall the definition of a \textit{realizable} abelian symmetry: an 
abelian symmetry is realizable if there are no larger abelian symmetry containing 
it.
Other approaches can be seen in Ref.\,\cite{schieren} and in the references in 
the introduction.
Note that Mathematica packages are available to promptly compute the 
SNF\,\cite{packages}.

\subsection{Example}
\label{sec:example}

Let us take as an example a Lagrangian depending on three scalar fields 
$\phi_1,\phi_2,\phi_3$, such as Higgs fields, appearing in the Lagrangian through 
only two phase sensitive terms, 
\eq{
\label{ex:terms}
\phi_2^\dag\phi_1\phi_2^\dag\phi_3+(\phi_3^\dag\phi_1)^2
\subset \lag\,.
}
We have suppressed the numerical coefficients.

The information on rephasing symmetries of the model is all encoded in the 
$D$-matrix\,\cite{abelian:yuk},
\eq{
\label{ex:D}
D=\mtrx{1&-2&1\cr2&0&-2}\,,
}
where each row corresponds to one phase sensitive term in the Lagrangian in
\eqref{ex:terms} by applying the following rule: attribute one unit of a
field specific charge to each field and list the cumulative charges 
of each field [we use the order $(\phi_1,\phi_2,\phi_3)$ in 
\eqref{ex:D}].

The appearance of the integer $D$-matrix can be understood as 
follows\,\cite{abelian:yuk}: apply general rephasing transformations
\eq{
\label{rephase}
\phi_j\to e^{i2\pi a_j}\phi_j\,.
}
Then the two terms in \eqref{ex:terms} gain phases 
$2\pi(a_1-2a_2+a_3)$ and $2\pi(2a_1-2a_3)$,
which should be integer multiples of $2\pi$ to leave the Lagrangian invariant.
In other words, the transformations \eqref{rephase} correspond to a rephasing 
symmetry if 
\eqali{
a_1-2a_2+a_3&=b_1 ~~\text{is integer}\,,\cr
2a_1-2a_3&=b_2~~\text{is integer}\,.
}
In matricial notation, the condition is simply
\eq{
  \label{cond}
DA=B ~~\in~~\ZZ^2\equiv	\ZZ\times\ZZ\,,
}
where $A=(a_1,a_2,a_3)^\tp$ and $B=(b_1,b_2)^\tp$.
However, not all solutions of \eqref{cond} for $A$ are relevant: if all $a_i$ 
are integers, the solution is trivial. We are only interested in nontrivial 
solutions where $A\not\in \ZZ^3$, which correspond to nontrivial discrete or
continuous transformations of the type \eqref{rephase}.

Deciding if Eq.\,\eqref{cond} possesses nontrivial solutions for $A$ is most easily 
accomplished if we can transform basis for $A,B$ (within integers) so that $D$ has 
diagonal form. Such a basis can be always found.
By one elementary row operation and two column operations, the $D$-matrix can be 
transformed to the Smith Normal Form:
\eq{
\label{ex:snf}
\snf(D)=\mtrx{1&0&0\cr0&4&0}\,.
}
If we apply the necessary row and column operations on the respective identity 
matrices, we obtain
\eq{
\label{ex:RC}
R=\mtrx{1&0\cr-2&1},~~
C=\mtrx{1&2&1\cr0&1&1\cr0&0&1}\,,
}
which performs the diagonalization:
\eq{\label{ex:RDC}
RDC=\snf(D).
}
Hence, condition \eqref{cond} can be rewritten as
\eq{
  \label{new.basis}
D'A'=B'~\in~\ZZ^2\,,
}
where $D'=\snf(D)$, $A'=C^{-1}A$ and $B'=RB$.
Since $R,C$ are unimodular, they univocally map integer vectors into integer 
vectors and nontrivial solutions in the new basis correspond to nontrivial 
solutions in the original basis.

In the new basis, it easy to see that the only nontrivial solutions are
$A'=t(0,1,0)^\tp$, with $t=1/4,2/4,3/4$, and 
$A'=t(0,0,1)^\tp$, with continuous $t$.
In the original basis, they correspond to
\eqali{
 \label{sol}
A&=t(2,1,0)^\tp,~~t=1/4,2/4,3/4\,,\cr
A&=t(1,1,1)^\tp,~~t\in \RR\,.
}
where the generating vectors correspond to the second and third columns of $C$, 
respectively.
In terms of rephasing transformations \eqref{rephase}, the solution in the first 
line of \eqref{sol} corresponds to a discrete $\ZZ_4$ symmetry whereas the second 
solution corresponds to a continuous $U(1)$ symmetry. They correspond to 
rephasing symmetries \eqref{rephase}:
\eqali{
 \label{sol:2}
\ZZ_4:(\phi_1,\phi_2,\phi_3)&\sim 
(e^{i2\pi \frac{2k}{4}},e^{i2\pi \frac{k}{4}},1)\,,~~~k=0,1,2,
\cr
U(1):(\phi_1,\phi_2,\phi_3)&\sim (e^{i2\pi t},e^{i2\pi t},e^{i2\pi t}),
~~t\in \RR.
}

We can now see that all information on abelian symmetries, i.e., the symmetries 
and their generators, can be extracted from $\snf(D)$ and $C$, respectively, 
without the need to perform all the basis change.
We read these informations in the following way:
the SNF form \eqref{ex:snf} is unique and implies that our symmetry is 
$\ZZ_4\times U(1)$, where $\ZZ_4$ is given by the nonzero and non-unit factor $4$ 
in the diagonal and $U(1)$ is characterized by the presence of one zero column.
The $U(1)$ symmetry is generated by the last column of $C$ in \eqref{ex:RC} since
\eq{
\label{ex:U1}
D\bss=\bs{0}, \text{ when } \bss=(111)^\tp\,.
}
On the other hand, the $\ZZ_4$ symmetry is generated by the second column of $C$ 
since
\eq{
\label{ex:Z4}
D\bss=4\mtrx{0\cr 1}=0\mod 4, \text{ when } \bss=(210)^\tp\,,
}
which means $\frac{1}{4}\bss$ is a nontrivial solution to \eqref{cond} associated 
to the nontrivial factor $4$ in $\snf(D)$.
For this reason, when we say $\bss$ is the generator of $\ZZ_d$, we adopt the 
convention that the corresponding rephasing transformation \eqref{rephase} uses the 
vector divided by the factor $d$ as $A=\frac{1}{d}\bss$.

\subsection{Charges are not unique}
\label{sec:charges}

The charges that generate each discrete symmetry are not unique because 
$R$ and $C$ in \eqref{ex:RDC} are not unique.
For example, the following $R,C$ also diagonalize $D$:
\eq{
R=\mtrx{1&0\cr 2&1},~
C=\mtrx{0&1&1\cr0&0&1\cr1&-1&1}.
}
Therefore the charge $\bss'=(1,0,-1)^\tp$ could be equally used instead of 
\eqref{ex:Z4} as a generator of $\ZZ_4$.
As a necessary condition, a charge $\bss$ should be a vector of relatively prime 
integers because $C$ is a invertible integer matrix which should have determinant 
$\pm 1$. If a column has a common factor $k$, $|k|\neq 1$, then the determinant is 
also divisible by $k$.

The first freedom for charges is calculation modulo $k$.
In our example, we often can make calculations in $\ZZ_4$, modulo 4. For example,
\eq{
\bss=3\mtrx{2\cr1\cr0}\mod 4=\mtrx{2\cr-1\cr0}\,,
}
can be equally used as charge instead of \eqref{ex:Z4} because $a^3$ is a 
generator of $\ZZ_4$ generated by $a$ (in multiplicative notation).
The vector 
\eq{
\bss=2\mtrx{2\cr1\cr0}\mod 4=\mtrx{0\cr 2\cr0}\,,
}
can not be used as a generator because it can be divided by $2$ and it only 
generates the $\ZZ_2$ subgroup of $\ZZ_4$, as $a^2$ in $\ZZ_4$ generated 
by $a$. One can check only $\ZZ_2$ is generated by following the action 
\eqref{rephase} with $A=\frac{1}{4}\bss$.

In the presence of one or more $U(1)$ symmetries, we can also add any 
combination of their charges at will. For example,
\eq{
\bss=\mtrx{2\cr1\cr0}\to 
\bss'=\mtrx{2\cr1\cr0}-\mtrx{1\cr1\cr1}=\mtrx{1\cr0\cr-1},
}
also works as a charge vector.
On the other hand, 
\eq{
\bss''=2\mtrx{2\cr1\cr0}-\mtrx{1\cr1\cr1}=\mtrx{3\cr1\cr-1},
}
leads to
\eq{
D\bss''=\mtrx{0\cr 8}\,.
}
This means $\bss''$ only generates $\ZZ_2$ instead of $\ZZ_4$ and cannot be used as 
a charge vector, although it consists of relatively prime integers.
One can see this more clearly by checking the transformation \eqref{rephase} 
with $A=\frac{1}{4}\bss''$, with the result being valid modulo the $U(1)$ symmetry.
(Note that $\ZZ_4$ would be still generated if we had used $A=\frac{1}{8}\bss''$ 
instead.)

\subsection{Notation}
\label{sec:notation}

In general, we can analyze the existence of discrete symmetries defined by $D$ in 
terms of the linear equation
\eq{
\label{system}
DA=B\,,
}
where $A\in\ZZ^{n_F}$ and $B\in\ZZ^{n_t}$, where $n_F$ and $n_t$ are the number of 
(complex) fields and the number of phase sensitive terms in the Lagrangian, 
respectively.
(Pairs of different hermitean conjugate terms are counted as one term.)
Compared to Ref.\,\cite{abelian:yuk}, we are already rescaling the 
equation \eqref{system} by $2\pi$; see \eqref{rephase}.

The existence of the SNF for $D$,
\eq{
\snf(D)=RDC=\diag(d_1,d_2,\cdots,d_n,0,\cdots,0)\,,
}
where each $d_i$ divides $d_{i+1}$,
allows us to diagonalize Eq.\,\eqref{system} into $D'A'=B'$, where $D'=\snf(D)$; 
cf. Eq.\,\eqref{new.basis}.
Given the SNF, we conclude that 
\eq{
D\bss^{(i)}=d_i\bw^{(i)}\,,
}
where $\bss^{(i)},\bw^{(i)}$ correspond to the $i$-th column of $C,R^{-1}$,
respectively. They are generators of the lattices $\ZZ^{n_F}$ and $\ZZ^{n_t}$, 
respectively.
Obviously, we get $D\bss^{(i)}=\bs{0}$, for $i>\rank D$.
The decomposition thus signals a symmetry 
$\ZZ_{d_1}\times\ZZ_{d_1}\times\cdots\times\ZZ_{d_{n}}\times [U(1)]^k$, where 
$k$ is the number of zero factors.
Each $\ZZ_{d_i}$ is generated by charge vector $\bss^{(i)}$ which is related to 
the actual rephasing transformation \eqref{rephase} with phases (divided by $2\pi$)
\eq{
 \label{phases}
A=\frac{1}{d_i}\bss^{(i)}\,.
}

In general, we denote the rows of $D$ by
\eq{
D=\mtrx{\bd^{(1)}\cr\bd^{(2)}\cr\vdots\cr \bd^{(n_t)}\cr}\,.
}
The rank of $D$ is the number of linearly independent (see Sec.\,\ref{sec:LI}) 
rows (or columns) of $D$ and it can be counted as the number of nonzero factors in 
the SNF of $D$.
Often, we will use matrices of full rank and denote its rank as $n$. For such a 
matrix, its SNF is 
\eq{
\snf(D)=
  \mtrx{d_1&&&&0\cdots\cr&d_2&&&0\cdots\cr&&\ddots&&\vdots&\cr&&&d_n&0\cdots}\,.
}

The following proposition will be useful.
\bigskip

\begin{minipage}{.9\textwidth}
\noindent\textbf{Proposition:} If, by adding a row $\bu$ to a matrix $D$ of $n$ 
rows, its SNF remains the same, except for an additional zero row, then $\bu$ is an 
integer linear combination of the rows of $D$.
The converse also holds.
\end{minipage}

\bigskip\noindent
We can see the converse is trivially valid because the last row $\bu$ can be 
removed by elementary row operations. 
The proof of the main proposition is given in appendix \ref{ap:teo}.

\subsection{Linear independence}
\label{sec:LI}

For real or complex vectors spaces (a vector space over any field) such as $\RR^n$ 
or $\CC^n$ we know that the following statements are equivalent: (i) a set 
$\{\bu_1,\bu_2,\cdots,\bu_k\}$ is linearly dependent; (ii) one of the vectors in 
the set $\{\bu_1,\bu_2,\cdots,\bu_k\}$ can be written as a linear combination of 
the rest of vectors.

Here we need to consider vectors of integer components and they live in $\ZZ^n$.
These spaces are similar to vector spaces (they are called modules) and most of the 
usual properties of vectors spaces remain. The equivalence of properties (i) and 
(ii), however, does not hold.
In this article, when we refer to linear independence, we will refer to such a 
property over the integers (in $\ZZ^n$) if not specified otherwise.

As an example, let us take three vectors in $\ZZ^3$:
\eq{
\bu_1=(2,-1,-1), ~
\bu_2=(-1,2,-1), ~
\bu_3=(2,-2,0).
}
These three vectors are linearly dependent since 
\eq{
2\bu_1+(-2)\bu_2+(-3)\bu_3=\bs{0}\,.
}
However, we cannot write any $\bu_i$ as an \textit{integer} linear combination of 
two other vectors.

The SNF can be used to extract these properties as follows:
\eqali{
\snf(\bu_1,\bu_2,\bu_3)&=\diag(1,1,0),~\cr
\snf(\bu_1,\bu_2)&=\diag(1,3),~\cr
\snf(\bu_1,\bu_3)&=\diag(1,2),
}
where $\snf(\bu_1,\bu_2)$ denote the Smith normal form of a matrix composed of 
rows $\bu_1,\bu_2$.
The linear dependence of three $\bu_i$ can be seen in the zero factor of the first 
equation: the rank is two.
The sets $\{\bu_1,\bu_2\}$ and $\{\bu_1,\bu_3\}$ are linearly independent but they 
generate different subspaces (lattices) of $\ZZ^3$ and their SNF are different.

\subsection{Height of vector}
\label{sec:type}

Take a vector (row) $\bu$ of a $D$-matrix derived from phase sensitive terms 
in a Lagrangian. We use the quantity ($l_1$-norm)
\eq{
h(\bu)=\sum_{i}|u_i|\,.
}
to classify vectors into different \textit{heights}. We will call $\bu$ as a 
height-$k$ vector when $h(\bu)=k$.
This quantity counts the number of phase sensitive fields in the 
Lagrangian term associated to $\bu$. 
For example, for a potential $V$ composed of scalar fields $\phi_i$, we can 
associate
\eqali{
\phi^\dag_1\phi_2\subset\lag &\implies \bu=(-1,1,0,\cdots)\,,\cr
\phi^\dag_1\phi_2\phi^\dag_1\phi_3\subset\lag &\implies 
\bu=(-2,1,1,0,\cdots)\,,\cr
\phi^\dag_1\phi_2\phi^\dag_3\phi_1\subset\lag &\implies 
\bu=(0,1,-1,0,\cdots)\,.\cr
}
In the first two examples, $h(\bu)$ corresponds to the number of fields whereas 
in the third example the term is not sensitive to the phase 
of $\phi_1$.
Elementary row operations does not maintain the height fixed but this 
classification will still be useful; see e.g. Sec.\,\ref{sec:discard:N>=6}.

\section{Matching up symmetries}
\label{sec:match}

From this point on, we will concentrate on the N-Higgs-doublet models with quarks,
which can be divided into two sectors: the Yukawa interactions and the Higgs 
potential. We denote the $D$-matrix for the terms in the Yukawa interactions and in 
the potential as $D^Y_\full$ and $D^V$, respectively.
Applications to the lepton sector or to other models can be considered in an 
analogous manner.
We only consider renormalizable terms and assume that discrete symmetries are valid 
up to very high energies (e.g., remnants of local gauge symmetries) so that no 
appreciable breaking is expected through nonrenormalizable terms. 
We consider the most general Yukawa interactions in the form\,\cite{abelian:yuk}
\eq{
- {\cal L}_Y = \Gamma^{(j_\phi)}_{j_L j_d}\bar Q_{Lj_L} \phi_{j_\phi} d_{Rj_d} + 
\Delta^{(j_\phi)}_{j_L j_u}\bar Q_{Lj_L} \tilde \phi_{j_\phi} u_{Rj_u} +
h.c.,
}
where $j_L,j_d,j_u$ run from 1 to 3 and $j_\phi$ runs from 1 to $N$.

The current knowledge is summarized for small $N$ in the following table containing 
the bounds for the order of realizable abelian symmetries in the potential 
($G_\phi$) and in the Yukawa interactions 
$(G_q)$:
\eq{
\label{TableGF}
\begin{array}{r|ccccc}
N & 2 & 3 & 4 & 5 & 6 \\\hline
|G_q| \le  & 3 & 5 & 8 & 12 & \underline{16} \\
|G_\phi| \le & 2 & 4 & 8 & 16 & 32
\end{array}
}
The upper bounds in the last row where shown in \cite{abelian:V}  and they
correspond to the exact upper boundary.
The upper bounds in the first row were proven in \cite{abelian:yuk} and all cases
up to $N=5$ were explicitly shown to be realizable (we mark the unproven case
underlined).
Algebraically, we have\,\cite{abelian:V,abelian:yuk}
\eqali{
  \label{bounds}
\boundY &=\left\{\begin{array}{ll}
    n^2/3\,, & \text{ if $n$ is divisible by $3$,}\cr
    (n^2-1)/3\,, & \text{ if $n$ is not divisible by $3$}
    \end{array}\right.
\cr
\boundV &=2^{N-1},
}
where $n=N+1$.
As before, $\boundV$ is the exact boundary while $\boundY$ is an upper bound.
We can see that the maximal group size grows much quickly for the 
potential than for the Yukawa interactions. Qualitatively, that happens because as 
$N$ grows the number of Lagrangian terms in the potential grows much quickly ($\sim 
N^4$) than the number of Yukawa terms with three families of quarks ($\sim N$).

We now seek conditions for a given abelian symmetry in one sector to be compatible 
with the symmetry in the other sector.
Let us denote the symmetries in the Yukawa sector and in the potential as $G_q$ and
$G_\phi$, respectively. Often these groups will refer to abstract groups and we 
disregard the automatic symmetries $U(1)_B,U(1)_Y$, and others that might appear. 
For example, if a field does not appear in a sector of the 
Lagrangian, then its rephasing is automatically a symmetry.
The true symmetry of the theory is $G_F={\tG_q\cap\tG_\phi}$.\,\footnote{
For this identification, we need to consider $\tG_q$ and $\tG_\phi$ as full 
subgroups inside $[U(1)]^{n_F}$, the rephasing symmetries of all $n_F$ complex 
fields involved.
Therefore, $\tG_q=G_q$ but $\tG_\phi$ contains the additional rephasing symmetries 
for the fermion fields besides $G_\phi$.
The distinction may be important in some situations; see remarks in appendix 
\ref{ap:remark}.
}
Moreover, $G_q\neq G_F$ or $G_\phi\neq G_F$ (not isomorphic) is naturally possible 
only if they arise accidentally from $G_F$ or a common smaller symmetry contained in 
$G_F$. We also note that in this specific setting (NHDMs), $G_F$ needs to be 
entirely broken after EWSB to allow nonzero and nondegenerate quark masses with 
mixing\,\cite{sb.quarks}.

To set the notation for applying the Smith normal form method,
we use the basis with ordering $(\phi_a;Q_{iL};d_{iR};u_{iR})$, corresponding to 
$n_F=N+9$ complex fields (columns of $D^Y_\full,D^V$).
However, the relevant information from $D^Y_\full$ on nontrivial symmetries may be 
extracted from a reduced matrix $D^Y$, which depends only on the $N+3$ fields 
$(\phi_a;Q_{iL})$.
A different notation was used in Ref.\,\cite{abelian:yuk} where the reduced matrix 
was denoted by $\tD$ whereas the full matrix was simply $D$.
We also note that, for $N\ge 6$, the rows of $D^Y$ can not be generic
but we can exclude cases by assuming generic matrices.
Now, we proceed to analyze the compatibility of $D^Y$ and $D^V$.

The minimal number of rows (phase sensitive terms in the Lagrangian) necessary to
sustain discrete symmetries in $D^Y$ and $D^V$ are $n=N+1$ and $n-2$, respectively.
To analyze the symmetry of the whole theory we also define the $D$-matrix
for the whole theory as
\eq{
D^T=\mtrx{D^Y\cr\hline D^V}\,.
}
If $D^Y$ and $D^V$ have $n$ and $n-2$ rows, respectively, then $D^T$ has $2n-2$
rows.

For the minimal number of rows, the generic structure of $D^Y$ and $D^V$ reads
\eq{
\label{tD:Y,V}
D^Y=\left(\begin{array}{c|c}
\bp_1 & \bq_1\\
\bp_2 & \bq_2\\
\vdots & \vdots \\
\bp_{n} & \bq_{n}
\end{array}\right)
\,,\quad
D^V=\left(\begin{array}{c|c}
\br_1 & \bs{0}_3\\
\br_2 & \bs{0}_3\\
\vdots & \vdots \\
\br_{n-2} & \bs{0}_3
\end{array}\right)\,.
}
The $\bp$-, $\bq$- and $\br$-vectors have length $n-1\,(N), 3$ and $n-1$, 
respectively.
The $\bp$-vectors are of the form $(1,-1,0,\ldots)$, or permutations, the
$\bq$-vectors are of the form $(1,-1,0)$, or any permutation, and the $\br$-vectors
are made up of one $\bp$-vector or the sum of two $\bp$-vectors.
The columns of zeros denoted by $\bs{0}_3=(0,0,0)$ correspond to three fields 
$Q_{iL}$ which do not appear in the potential.
If we use the classification of Sec.\,\ref{sec:type}, $\bp$- and $\bq$-vectors are 
of height 2 (one of them can be of height 0) and $\br$-vector are of heights 2 or 4.
Given the automatic conservation of $U(1)_Y$ and $U(1)_B$, when we sum $\bp$- or 
$\bq$-vectors, their height always change in units of two. For example, by summing 
two $\bq$-vectors, we can obtain vectors of heights $0,2$ or $4$.

We start by analyzing the conditions for \textit{faithful compatibility}, i.e., 
$G_q\simeq G_\phi$ (isomorphic). We say $G_q$ (Yukawa) and $G_\phi$ (potential) are 
faithfully compatible symmetries in the whole theory if all the integers that appear 
in the SNF of $D^Y,D^V$ and $D^T$ are the same:
\eqali{
\label{compatible}
\text{(I.Y)}\quad& \snf(D^Y)=
\mtrx{d_1&&&&0&0\cr&d_2&&&0&0\cr&&\ddots&&\vdots&\vdots\cr&&&d_n&0&0}\,,
\cr
\text{(I.V)}\quad& \snf(D^V)=
\mtrx{d_3&&&&\bs{0}_4\cr&&\ddots&&\vdots\cr&&&d_n&\bs{0}_4}\,,
\cr
\text{(II)}\quad& \snf(D^T)=
\mtrx{d_1&&&&0&0\cr&d_2&&&0&0\cr&&\ddots&&\vdots&\vdots\cr&&&d_n&0&0\cr
0&\cdots&&&0&0\cr \vdots &\cdots&&&\vdots&\vdots\cr 0&\cdots&&&0&0}\,.
}
Often most of the factors $d_i$ will be unity. In particular, we
must necessarily have $d_1=d_2=1$.
Generalization to theories with more sectors can be done by considering the 
separation of the theory into one sector and the rest.

If the factors appearing in (I.Y), (I.V) and (II) do not all match but (II) still 
contains non-unit integers, we still say $G_q$ and $G_\phi$ are compatible and the 
symmetry of the whole theory ($G_F$) is different from the symmetry of one or both 
of its subsectors ($G_q$ or $G_\phi$).
In any of these cases, $G_F$ is the true symmetry of the theory if there are no
additional $G_F$-symmetric terms that reduces $G_q$ or $G_\phi$ (and possibly 
$G_F$).
If $G_q$ or $G_\phi$ changes after the addition of $G_F$-symmetric terms, we say 
they are \textit{not compatible}.
In practice, we can begin with some $G_\phi$ or $G_q$ and start to add 
$G_F$-symmetric terms if possible. If the process stops with some finite $G_F$, 
then the groups $G_q$ and $G_\phi$ in the last step are compatible. Of course, the 
process may continue until we have no symmetry.
Natural compatibility that is not faithful may happen in two situations: either 
the symmetry in some subsector contains accidental symmetries not present in 
$G_F$ or only a subgroup of $G_F$ is faithfully represented in some subsector.
One example of the former case can be seen in Sec.\,\ref{sec:V->Y} where $G_F\simeq 
G_\phi=\ZZ_8$ but $G_q=U(1)\supset G_F$. 
Appendix \ref{ap:remark} contains an example for the second case.
In general settings, not faithful natural compatibility is difficult to characterize 
and thus we will concentrate on the faithfully compatible cases.

Let us return to the conditions in \eqref{compatible} and analyze its consequences.
Such conditions imply that the rank of $D^T$ are the same
as the rank of $D^Y$ and we can conclude that the $\br$-vectors in $D^V$ should
be particular linear combinations of the $\bp$-vectors of $D^Y$, constrained by
the condition that the contribution from the $\bq$-vectors should vanish.
Let us assume, without loss of generality, that $\{\bq_1,\bq_2\}$ are linearly
independent.
By row operations we can transform $D^Y$ in \eqref{tD:Y,V} to
\eq{
\label{tD:Y:2}
D^Y\sim\left(\begin{array}{c|c}
\bp_1 & \bq_1\\
\bp_2 & \bq_2\\
\bp_3' & \bs{0}_3\\
\vdots & \vdots \\
\bp_{n}' & \bs{0}_3
\end{array}\right)
\,.
}
From the structure of $\bq$-vectors, we conclude that the $\bp_i'$ vectors are made
of \textit{at most three} $\bp$-vectors and it should be of height at most 6.
Full compatibility and the proposition of Sec.\,\ref{sec:notation} then requires 
that $\bp_i'$ be integer linear combinations of
$\br_i$ and the reverse should be also true.
Therefore we conclude that the following matrices should be equivalent by row
operations for faithful compatibility:
\eq{
\label{condition:full}
\mtrx{\bp_3'\cr \vdots\cr \bp_n'}\sim \mtrx{\br_1\cr \vdots\cr \br_{n-2}}\,.
}

\section{Maximal symmetries for three Higgs doublets}
\label{sec:max:3h}

Before considering the three-Higgs-doublet case, let us review the simpler 2HDM
with $\ZZ_2$ symmetry.
We know that the simple $\ZZ_2$ symmetry can be implemented in the 2HDM to
naturally suppress flavor changing neutral currents\,\cite{nfc}, and it is
often used as a benchmark model for testing 2HDMs\,\cite{branco:review}.
That implementation, actually, leads to an accidental Peccei-Quinn symmetry
in the Yukawa sector which is only broken to $\ZZ_2$ in the Higgs potential through
the term $(\phi_1^\dag\phi_2)^2$.
(See end of Sec.IV.B in Ref.\,\cite{joao:abelian}.)
A true 2HDM with $\ZZ_2$ symmetry\,\footnote{An accidental CP symmetry is always 
present in the potential\,\cite{2hdm:sym}.}
in both Yukawa sector and Higgs potential can be
constructed\,\cite{joao:abelian} but it requires flavor dependent $\ZZ_2$ charges
and thus to potentially harmful FCNC interactions unless the non-SM scalar bosons
are very heavy.

Returning to three-Higgs-doublets, $N=3$, we can see in Table \eqref{TableGF} that
$|G_q|_{\rm max}=5\ge|G_\phi|_{\rm max}=4$. Thus the abelian symmetries
$G_F$ which are realizable for 3HDM should have order equal or less than 4.
We show here that $|G_F|=4$ is realizable and we explicitly give the backbone 
structure from which the whole theory can be reconstructed\,\cite{abelian:yuk}.
Let us recall that the $D$-matrix for Yukawa interactions in
this case has the form \eqref{tD:Y,V},
\eq{
\label{DY:3h}
D^Y=\left(\begin{array}{c|c}
  \bp_1 & \bq_1\cr
  \bp_2 & \bq_2\cr
  \bp_3 & \bq_3\cr
  \bp_4 & \bq_4
  \end{array}\right)\,.
}
The $\bp$-vectors ($\bq$-vectors) correspond to the fields $\phi$ ($Q_L$) and
have the form $(1,-1,0)$ or similar with permuted entries. There are 6
possibilities with only three nonparallel vectors, which we denote as being of
distinct type:
\eq{
\label{q-vectors}
\bq_i\in\{\pm(1,-1,0),\pm(0,1,-1),\pm(1,0,-1)\}\,.
}
Only two of them are linearly independent.
We have seen in \cite{abelian:yuk} that $|G_q|_{\rm max}=5$ is achieved only when 
the $\bq$-vectors are picked most evenly among the different types, i.e., the
number $n_i$ of a type $i$ vector should be $(n_1n_2n_3)=(211)$ and the order of
the group is bounded by $m=n_1n_2+n_2n_3+n_3n_1=5$.
To get $|G_q|=4$, we can allow $(n_1n_2n_3)=(211)$ or $(n_1n_2n_3)=(220)$.
Since $\bp$-vectors and $\bq$-vectors are the same in this case, the same 
distribution of different types also applies to the $\bp$-vectors.

By row operations, we can obtain the form \eqref{tD:Y:2} which, in this case, reads
\eq{
\label{DY:3h:2}
D^Y\sim \left(\begin{array}{c|c}
  \bp_1 & \bq_1\cr
  \bp_2 & \bq_2\cr
  \bp_3' & \bs{0}\cr
  \bp_4' & \bs{0}
  \end{array}\right)\,,
}
where $\bp'_i$ are at most height-6 vectors, as they are integer combinations of
$\bp_i$ and $\bp_1,\bp_2$ (we are assuming $\bq_1,\bq_2$ nonparallel).
When calculating the SNF form of $D^Y$, we can see the nontrivial integers come
from the submatrix $(\bp_3',\bp_4')$.
Analogously, the reduced $D$-matrix of the potential is
\eq{
D^V=\left(\begin{array}{c|c}
  \br_1 & \bs{0}\cr
  \br_2 & \bs{0}
  \end{array}\right)\,.
}
We know that the vectors $\br_1,\br_2$ are at most height-4 vectors.
The compatibility condition in \eqref{condition:full} requires that
$\bp_3',\bp_4'$ be integer linear combinations of $\br_1,\br_2$ and vice-versa.
Let us see if that is possible for $|G_F|=4$.

To satisfy the conditions (I) of \eqref{compatible}, the SNFs of the sectors
$(\br_1,\br_2)$, $(\bp_3',\bp_4')$ should contain the same non-unit integers that
multiply to 4. The only possibilities are one unit integer
$4$ or two non-unit integers $(2,2)$.
These possibilities correspond to the symmetries $\ZZ_4$ and $\ZZ_2\times\ZZ_2$,
respectively.
In the potential, these possibilities correspond to the backbone
structures\,\cite{abelian:yuk}
\eqali{
\label{backbone:V:4}
\ZZ_4&:~~D^V=
\left(\begin{array}{ccc|c}
  2&-1&-1 & \bs{0}\cr
  0&2&-2 & \bs{0}
  \end{array}\right)\,,
\cr
\ZZ_2\times\ZZ_2&:~~D^V=
\left(\begin{array}{ccc|c}
  2&-2&0 & \bs{0}\cr
  0&2&-2 & \bs{0}
  \end{array}\right)\,.
}
Other possibilities amount to relabeling of Higgs doublets.
These structures lead to the charges
\eqali{
\label{Z4:charges}
\ZZ_4&:~~\bss^V=(1,2,0;***)^\tp\sim(0,1,-1;***)^\tp\,,\cr
\ZZ_2\times\ZZ_2&:~~\bss^V=(1,0,0;***)^\tp\otimes(0,0,1;***)^\tp\,,
}
where we marked the undefined $Q_{iL}$ charges by $*$ and subtracted one unit
of the conserved charge $(111;\bs{0})$ on $\bss^V$ of $\ZZ_4$.
Notice that we use $\bss$ instead of $\tbss$ when compared to 
Ref.\,\cite{abelian:yuk}.

Let us now analyze if $\bp_3',\bp_4'$ can be of height-6.
All height-6 vectors in this case are $(3,-2,-1)$, $(3,-3,0)$ or vectors with
permuted entries.
By applying a height-6 vector (whose sum of components are zero) to the
$\ZZ_4$-charge in \eqref{Z4:charges}, it is not possible to obtain zero or a
multiple of 4. Analogously, for $\ZZ_2\times\ZZ_2$-charge it is not possible to
obtain a pair $(a_1,a_2)$ where each of $a_i$ is zero or multiple of 2.
Therefore $\bp_3',\bp_4'$ should be of height-4 or less. But from condition 
(I.Y) of \eqref{compatible} neither of them can be of height-2 (it leads to unit 
factors in SNF). Thus we conclude they must be both of height-4.
Therefore, $(\bp_3',\bp_4')$ should be also of the form \eqref{backbone:V:4}.

Now, we can check that the only height-4 vectors that give zero or multiples of 4 
after applying the $\ZZ_4$-charge are the two rows of $D^V$ ($\ZZ_4$) in
\eqref{backbone:V:4}. 
For $\ZZ_2\times\ZZ_2$, $\bp_3',\bp_4'$ should correspond to integer
combinations of $\br_i$. Then $\bp_3',\bp_4'$ are two nonparallel vectors of the
form $(2,-2,0)$ or permuted entries.
In either case, given that the SNF form of $(\bp'_3,\bp_4')$ should also
correspond to a finite symmetry, we conclude that $\bp'_3,\bp_4'$ must correspond
to the rows of $D^V$ in \eqref{backbone:V:4}, except for a possible
interchange or overall minus sign. 

The following step would be to list all $D^Y$ matrices \eqref{DY:3h} that are
compatible with \eqref{DY:3h:2}, where $(\bp_3',\bp_4')$ are of the form 
\eqref{backbone:V:4}. Since this backward analysis would be lengthy, we employ a
different strategy in appendix \ref{ap:order4-3hdm} to characterize all possible
matrices $D^Y$ of the form \eqref{DY:3h} which are compatible with $D^V$ in
\eqref{backbone:V:4}.
We find \textit{only one backbone structure for both symmetries $\ZZ_4$ and
$\ZZ_2\times\ZZ_2$}.
We list them in the following together with their complete charges.
By using charges, Yukawa textures can be recovered from the recipe given in 
Sec.\,6.2 of Ref.\,\cite{abelian:yuk}.
After applying the recipe, we need to check if the theory realizes the discrete 
symmetry or if it is invariant by a larger  $U(1)$ symmetry containing the
discrete symmetry as a subgroup. The former is true if $D^Y\bss \neq \bs{0}$.

\subsection{$\ZZ_4$-3HDM}

The unique backbone structure for $\ZZ_4$-3HDM is 
\eq{
\label{backbone:z4-3h}
D^Y=
\left(
\begin{array}{ccc|ccc}
 1 & -1 & 0 & 1 & -1 & 0 \\
 1 & -1 & 0 & -1 & 1 & 0 \\
 0 & 1 & -1 & 0 & 1 & -1 \\
 1 & 0 & -1 & 0 & -1 & 1 \\
 0 & 1 & -1 & -1 & 0 & 1 \\
 1 & 0 & -1 & 1 & 0 & -1
\end{array}
\right),\quad
D^V=
\left(
\begin{array}{cccccc}
 2 & -2 & 0  \\
 1 & 1 & -2 
\end{array}
\right)\,,
}
where we reordered the matrix in \eqref{backbone:V:4} for compatibility and 
suppressed the automatic zeros in $D^V$ for simplicity.
The details can be found in appendix \ref{ap:order4-3hdm}.
We should note that $D^Y$ can have 4 rows to sustain a finite symmetry but we
discover that the possible backbone structures for $D^Y$ which contains 4 rows are
all compatible with the structure above. 
Different possibilities only amounts to reordering $\phi_a$ and $Q_{iL}$ fields.
One can also check that if we write $D^Y$ in the form \eqref{tD:Y:2}, we obtain
exactly $D^V$ in the lower rows.

The $\ZZ_4$-charges arising from \eqref{backbone:z4-3h} reads
\eq{
\label{3h-z4:charge}
\bss_{\ZZ_4}=(1,-1,0;-1,1,0)\,.
}
We can see these charges in the $\phi$-sector are analogous to \eqref{Z4:charges}
after reordering.

\subsection{$\ZZ_2\times\ZZ_2$-3HDM}

There is only one backbone structure for $\ZZ_2\times\ZZ_2$-3HDM:
\eq{
\label{backbone:z2z2-3h}
D^Y=
\left(
\begin{array}{ccc|ccc}
 1 & -1 & 0 & 1 & -1 & 0 \\
 1 & -1 & 0 & -1 & 1 & 0 \\
 0 & 1 & -1 & 0 & 1 & -1 \\
 0 & 1 & -1 & 0 & -1 & 1 \\
 1 & 0 & -1 & -1 & 0 & 1 \\
 1 & 0 & -1 & 1 & 0 & -1
\end{array}
\right)
,\quad
D^V=
\left(
\begin{array}{cccccc}
 2 & -2 & 0  \\
 0 & 2 & -2 \\
 2 & 0 & -2 
\end{array}
\right)\,.
}
The corresponding charges are
\eq{
\label{3h-z2z2:charge}
\bss_{\ZZ_2\times\ZZ_2}=(1,0,0;1,0,0)\otimes(0,0,1;0,0,1)\,.
}
One can also check that if we write $D^Y$ in the form \eqref{tD:Y:2}, we obtain
exactly $D^V$ in the lower rows.

\section{Discarding maximal symmetries}
\label{sec:discard}

We show here that the abelian symmetry in the table of \eqref{TableGF},
with smallest order for each $N$, is not compatible in both the Yukawa sector and
the Higgs potential for $N\ge 4$, and its imposition leads to a continuous symmetry 
in the potential.

We see in the table of \eqref{TableGF} that $\boundV\ge\boundY$ for $N\ge 4$.
The maximal order of the groups allows, in principle, a common abelian group
$G_F=G_\phi=G_q$ with order $|G_F|=|G_q|_{\max}\le \boundY$.
We will show that $G_\phi$ compatible with $G_q|_{\rm max}$ can not be found.

\subsection{Discarding maximal symmetries for $N=4$ and $N=5$}
Let us take $N=4$ and $N=5$ explicitly.
We begin with $N=4$ where we investigate if $\ZZ_8$ symmetry is realizable in the
whole 4HDM theory.
From the backbone structure of the Yukawa interactions of $\ZZ_8$-4HDM in
Ref.\,\cite{abelian:yuk},
\eq{
\label{4h:Y:Z8}
D^Y_{N=4}(\ZZ_8) = 
\left(
\begin{array}{cccc|ccc}
 0 & 1 & -1 & 0 & 1 & 0 & -1\\
 0 & 1 & 0 & -1 & -1 & 1 & 0\\
 1 & -1 & 0 & 0 & 0 & 1 & -1\\
 -1 & 0 & 1 & 0 & 1 & 0 & -1\\
 -1 & 0 & 0 & 1 & -1 & 1 & 0
\end{array}
\right)\,.
}
we can find the following $\ZZ_8$ charges for $(\phi;Q_L)$ fields:
\eq{
\label{charges:Z8}
\bss_{\ZZ_8}=(-1,1,4,0;1,0,-2)^\tp\,,
}
Note that we use a different convention from \cite{abelian:yuk} where we can
find (without the prime)
\eq{
\bss_{\ZZ_8}'=(0,2,1,-3;-1,2,0)^\tp\,.
}
We note that
\eq{
\bss_{\ZZ_8}-\bss_{\ZZ_8}'=(-1,-1,3,3;2,-2,-2)^\tp\sim (0,0,4,4;4,0,0)^\tp\,,
}
where in the last equivalence we have used invariance by $U(1)_Y$ and $U(1)_B$.
One can check that
\eq{
  D^Y_{N=4}(\ZZ_8)(\bss_{\ZZ_8}-\bss_{\ZZ_8}')=(0,-8,0,8,0)^\tp\,,
}
i.e., the zero vector if the components are taken modulo 8.

Now, analyzing the $\phi$ charges of \eqref{charges:Z8}, we can immediately see
that there is no quadratic invariants of the form $\phi^\dag_i\phi_j$ and the only
quartic invariant terms are
\eq{
\label{V:4h:Z8}
(\phi_4^\dag\phi_3)^2,~
\phi_1^\dag\phi_4\phi_2^\dag\phi_4,~\phi_1^\dag\phi_3\phi_2^\dag\phi_3\,.
}
These terms lead to the $D$-matrix:
\eq{
D^V=
\left(
\begin{array}{ccccccc}
 -1 & -1 & 2 & 0  \\
 -1 & -1 & 0 & 2  \\
 0 & 0 & 2 & -2  \\
\end{array}
\right)\,.
}
However, we can immediately see that the third row is the first row subtracted 
from the second and thus $D^V$ has rank two.
We need at least another linearly independent row to obtain a finite symmetry.
Thus the Higgs potential has an $U(1)$ symmetry.

Let us now turn to $N=5$. The backbone structure of the
Yukawa interactions for $\ZZ_2\times\ZZ_6$-5HDM is\,\cite{abelian:yuk}
\eq{
\label{backbone:5HDM-12}
D^Y_{N=5}(\ZZ_2\times\ZZ_6)
=
\left(
\begin{array}{ccccc|ccc}
 1 & -1 & 0 & 0 & 0 & 1 & 0 & -1\\
 1 & 0 & -1 & 0 & 0 & -1 & 1 & 0\\
 0 & 0 & 0 & 1 & -1 & 0 & 1 & -1\\
 0 & 1 & 0 & -1 & 0 & 1 & 0 & -1\\
 0 & 0 & 1 & -1 & 0 & -1 & 1 & 0\\
 -1 & 0 & 0 & 0 & 1 & 0 & 1 & -1
\end{array}
\right)\,.
}
The $\ZZ_2\times\ZZ_6$ charges can be found to be
\eq{
\label{charges:Z2Z6}
\bss_{\ZZ_2\times\ZZ_6}=(0,1,1,0,0;1,0,0)^\tp\otimes(1,3,0,-1,0;2,1,0)^\tp\,,
}
where the charges refer to $\ZZ_2$ and $\ZZ_6$ respectively.
We can check that the only quadratic term $\phi_i^\dag\phi_j$ invariant by $\ZZ_6$
is $\phi_3^\dag\phi_5$, but this term is not invariant by $\ZZ_2$.
Among the quartic terms, we can find the $\ZZ_6$ invariants
\eqali{
\label{V:5h:Z6}
\phi_i^\dag\phi_j\phi_k^\dag\phi_l,&~~i,j,k,l=3,5,\cr
\phi_2^\dag\phi_i\phi_2^\dag\phi_j,&~~i,j=3,5,\cr
\phi_1^\dag\phi_i\phi_4^\dag\phi_j,&~~i,j=3,5,\cr
\phi_1^\dag\phi_4\phi_2^\dag\phi_4&.
}
Among them, the $\ZZ_2$ invariants are
\eqali{
\label{V:5h:Z2Z6}
(ijkl)&=(3535),\cr
(ijkl)&=(2323),(2525),\cr
(ijkl)&=(1343),(1545),
}
where we use the shorthand $\phi_i^\dag\phi_j\phi_k^\dag\phi_l\sim(ijkl)$.
The $D$-matrix for these terms is
\eq{
D^{V}=
\left(
\begin{array}{ccccc}
 0 & 0 & -2 & 0 & 2 \\
 0 & -2 & 2 & 0 & 0 \\
 0 & -2 & 0 & 0 & 2 \\
 1 & 0 & -2 & 1 & 0 \\
 1 & 0 & 0 & 1 & -2
\end{array}
\right)\,.
}
We can see by inspection that two rows can be eliminated by row reduction
leading to a rank-3 matrix and hence the potential is $U(1)$ symmetric.

It is surprising that $\ZZ_8$ and $\ZZ_2\times\ZZ_6$ are proven to
be realizable symmetries in the potential of 4HDM and 5HDM, 
respectively\,\cite{abelian:V}.
However, we have seen here that the maximal symmetries $\ZZ_8$ and
$\ZZ_2\times\ZZ_6$ that are realizable in the Yukawa sector of 4HDM and 5HDM,
respectively, can not be extended to the respective potentials: 
they lead to a continuous symmetry and to Goldstone bosons in the Higgs
spectrum at the classical level.
The converse also occurs: the realization of $\ZZ_8$ symmetry in the 4HDM 
potential also leads to a continuous symmetry in the Yukawa sector of 4HDM;
see example in Sec.\,\ref{sec:V->Y}.
Analogous considerations apply to 5HDM.

We should also remark that when $G_q$ is discrete but $G_\phi$ is accidentally 
continuous, quantum corrections will generally turn the additional 
Goldstone bosons into pseudo-Goldstone bosons. We will not focus on this 
possibility here.

\subsection{Discarding maximal symmetries for $N\ge 6$}
\label{sec:discard:N>=6}

Here, we extend our results of the previous section and discard all $G_F=G_q|_{\rm 
bound}$ in table \eqref{TableGF} as realizable symmetries for NHDMs for $N\ge 6$.
The order for maximal $G_q$ is given in Eq.\,\eqref{bounds}.
There might be symmetries in the potential $G_\phi$ of the 
same order $|G_\phi|=\boundY$, for $N\ge 4$\,\cite{abelian:V}, but we show 
here that they can not be compatible.

We start by reconsidering the cases $N=4$ and $N=5$ from a different point of view. 
We then generalize this method to the cases $N\ge 6$.
Let us transform the backbone structures for the Yukawa interactions of 
$\ZZ_8$-4HDM and $\ZZ_2\times\ZZ_6$-5HDM to the form \eqref{tD:Y:2}:
\eq{
D_{N=4}(\ZZ_8) \sim
\left(
\begin{array}{cccc|ccc}
 0 & 1 & -1 & 0 & 1 & 0 & -1\\
 0 & 1 & 0 & -1 & -1 & 1 & 0\\
 1 & -3 & 1 & 1 & 0 & 0 & 0\\
 -1 & -1 & 2 & 0 & 0 & 0 & 0\\
 -1 & -1 & 0 & 2 & 0 & 0 & 0
\end{array}
\right)\,,
}
\eq{
D_{N=5}(\ZZ_2\times\ZZ_6)
\sim
\left(
\begin{array}{ccccc|ccc}
 1 & -1 & 0 & 0 & 0 & 1 & 0 & -1\\
 1 & 0 & -1 & 0 & 0 & -1 & 1 & 0\\
 -2 & 1 & 1 & 1 & -1 & 0 & 0 & 0\\
 -1 & 2 & 0 & -1 & 0 & 0 & 0 & 0\\
 -1 & 0 & 2 & -1 & 0 & 0 & 0 & 0\\
 -1 & 0 & 0 & -1 & 2 & 0 & 0 & 0
\end{array}
\right)\,.
}
We can see that the vector in the third row, $\bp_3'$ is a height-6 vector
 (sum of \textit{three} $\bp$-vectors), in contrast to the vectors $\bp'_i$, 
$i\ge 4$, which are height-4 vectors (sum of \textit{two} $\bp$-vectors).
The condition for faithful compatibility \eqref{condition:full} requires that $D^V$ 
has rows $\br_i$ that can be written as linear combination of vectors 
$\{\bp_3',\cdots,\bp'_n\}$. So there should exist a sequence of row operations on 
the set $\{\bp'_i\}$ that transform it to the set $\{\br_i\}$.
However, we can see by inspection that there are no elementary row operations that 
can transform $\bp_3'$ into a height-4 or height-2 vector. Therefore, there are no 
$D^V$ which is compatible with maximal $D^Y$ for $N=4$ and $N=5$.

Now we generalize the previous method to general $N\ge 6$ (the arguments are valid 
for $N\ge 4$).
For maximal $D^Y$, we need that $\bq_i$ be distributed into the three types 
\eqref{q-vectors} as evenly as possible. This is achieved if we assume 
$\bq_1,\bq_2,\bq_3$ are each of a different type and repeat the subsequent vectors 
as
\eq{
\{\bq_i\}=\{\bq_1,\bq_2,\bq_3,\bq_1,\bq_2,\bq_3,\bq_1,\cdots\}\,.
}
Then, we can write $D^Y$ in the form \eqref{tD:Y:2} by subtracting 
\eqali{
  \label{pi'}
\bp'_{3k+1}&=\bp_{3k+1}-\bp_1,~~ k\ge 1,\cr
\bp'_{3k+2}&=\bp_{3k+2}-\bp_2,~~ k\ge 1,\cr
\bp'_{3k}&=\bp_{3k}-\bp_3,~~ k\ge 2,
}
which define all $\bp'_i$ for $i\ge 4$.
The remaining $\bp_3'$ is obtained as
\eq{
\label{p3'}
\bp_3'=\bp_3-\bp_1-\bp_2\,,
}
by using the convention that $\bq_3=\bq_1+\bq_2$.
We prove the inexistence of $D^V$ which is compatible with maximal $D^Y$ in 
two steps: 
(a) we show that the $\bp'_i$ vectors in \eqref{pi'}, $i\ge 4$, are height-4 vectors 
whereas $\bp'_3$ in \eqref{p3'} has height 6;
(b) we demonstrate that the height of $\bp'_3$ can not be reduced by summing or 
subtracting any $\bp'_i$, $i\ge 4$, and thus the set $\{\bp_i'\}$, $i\ge 3$, cannot 
be transformed to the set $\{\br_i\}$ by row operations.
Step (a) is shown in appendix \ref{ap:pi'} while step (b) is detailed in appendix 
\ref{ap:p3'}.

One remark is in order.
As explained in Ref.\,\cite{abelian:yuk}, constructing $D^Y_{\rm full}$ for the full 
theory from our reduced $D^Y$ may not be feasible for $N\ge 6$.
However, such a difficulty is not relevant for the maximal group of order $\boundY$ 
which were shown here to be non-realizable as a discrete symmetry in both sectors.
The above difficulty might be relevant for nonmaximal groups of order less than 
$\boundY$.

\section{Building symmetric models}
\label{sec:extension}

The analysis of realizable abelian symmetries of NHDMs restricted to the Yukawa 
sector and to the Higgs potential separately were given respectively in 
Refs.\,\cite{abelian:yuk,abelian:V}.
In the previous sections, we have seen how to \textit{analyze} the compatibility 
between different sectors and have concluded that symmetries of order 
$\boundY$ in \eqref{bounds} are not realizable in the whole NHDM theory.
Here, we discuss methods (A) to construct the maximal number of allowed terms in 
the Lagrangian from the minimal number of terms, (B) to transfer the symmetry of 
the Yukawa sector to the potential, (C) to extend the symmetry of the 
potential to the Yukawa sector and, finally, (D) to construct the full 
Yukawa terms from the backbone structure.

\subsection{From minimal to maximal number of terms}
\label{min->max}

In the SNF method, each row in the extracted $D$-matrix corresponds to a Lagrangian 
term or a pair including the respective hermitean conjugate.
From the $D$-matrix, we can straightforwardly \textit{analyze} the abelian symmetry 
of the theory.
Often, we want to do the opposite: how can we \textit{construct} the  Lagrangian 
for a given realizable symmetry?
In general, an abelian symmetry can be sustained by a minimal number of terms 
in the Lagrangian and one of the problems is how we extend the minimal terms to 
the maximal terms allowed by symmetry.
The usual method is to take the charges (generators) and explicitly write down 
all possible terms. Here we present an alternative method of extending a $D$-matrix 
with the minimal number of rows to a $D$-matrix with the maximal number of 
compatible rows.
This method leads to a unique matrix if the rows of the $D$-matrix (Lagrangian 
terms) are restricted to certain types and the fields involved are fixed.

We illustrate the method by using one example.
Let us take the backbone structure for $\ZZ_2\times\ZZ_2$-3HDM in 
\eqref{backbone:z2z2-3h} and eliminate some rows until the minimum number of 4 and 
2, respectively:
\eq{
\label{example}
D^Y=
\left(
\begin{array}{ccc|ccc}
 1 & -1 & 0 & 1 & -1 & 0 \\
 1 & -1 & 0 & -1 & 1 & 0 \\
 0 & 1 & -1 & 0 & 1 & -1 \\
 0 & 1 & -1 & 0 & -1 & 1 \\
\end{array}
\right)
,\quad
D^V=
\left(
\begin{array}{cccccc}
 2 & -2 & 0  \\
 0 & 2 & -2 \\
\end{array}
\right)\,.
}
We can reobtain the matrices in \eqref{backbone:z2z2-3h} by using the following 
recipe:
take all integer linear combinations of the rows that match similar forms.
The proposition in Sec.\,\ref{sec:notation} ensures that this procedure is the 
necessary and sufficient condition to maintain the symmetry.
In our example, the rows of $D^Y$ should be of the form $(\bp_i,\bq_i)$ where both 
$\bp_i$ and $\bq_i$ are of the form $(1,-1,0)$ or permuted entries, whereas the 
rows of $D^V$ should have at most height 4 and its components should add to zero.
The only linear combinations of the rows of \eqref{example} that match the criteria 
are the ones given in \eqref{backbone:z2z2-3h}.
The rows in \eqref{backbone:z2z2-3h} would give the maximum number of terms in the 
Lagrangian compatible with the symmetry if their row correspond directly to 
Lagrangian terms. This is not the case for $D^Y$ which is a reduced $D$-matrix.
This method of extending the rows of $D$ is quite general and can be applied to any 
case where a $D$-matrix contains the minimum number of linearly independent rows.

\subsection{From the Yukawa sector to the potential}

We illustrate here a method to transfer the symmetry of the Yukawa terms of NHDMs 
to the Higgs potential. This method can be adapted to more general cases where the 
symmetry structure of a sector involving more fields needs to be extended to 
sectors with less fields, a typical example of the latter being the scalar 
potential. In this case, the extended $D$-matrix is unique, except for possible 
ways of rewriting.

We take the $\ZZ_2\times\ZZ_2$-3HDM again, given in \eqref{example}.
Suppose we only know the symmetry in the Yukawa sector as in $D^Y$ and we want to 
extend the symmetry to the Higgs potential.
To that end, transform $D^Y$ to the form \eqref{tD:Y:2}, i.e., eliminate the 
entries that does not correspond to Higgs fields:
\eq{
\label{example}
D^Y=
\left(
\begin{array}{ccc|ccc}
 1 & -1 & 0 & 1 & -1 & 0 \\
 0 & 1 & -1 & 0 & 1 & -1 \\
\hline
 2 & -2 & 0 & 0 & 0 & 0 \\
 0 & 2 & -2 & 0 & 0 & 0 \\
\end{array}\right)\,.
}
Take the rows with entries in $Q_{iL}$ eliminated (below the horizontal line) and 
consider it as $D^V$.
If $D^V$ is sufficient to sustain the discrete symmetry, then we have extracted a 
faithfully compatible $D^V$. This is the case here where each sector sustains a 
compatible $\ZZ_2\times\ZZ_2$ symmetry, which is the symmetry of the whole theory.
Even if the extracted $D^V$ is not enough to sustain the discrete symmetry, the 
symmetry in $D^V$ would still be compatible with $D^Y$ but additional checks are 
necessary to see if an accidental $U(1)$ arises after considering all compatible 
terms.

\subsection{From the potential to the Yukawa sector}
\label{sec:V->Y}

We cover here the remaining case where a discrete symmetry is present in one sector 
of a theory but it leads to an accidentally continuous symmetry in another sector.
In this case the symmetries in the two sectors are compatible but not faithfully
compatible as in \eqref{compatible}.
One prime example is the following: a discrete symmetry in the scalar 
potential but an accidental continuous symmetry in the Yukawa sector.
Here we give an example of how to construct theories with such a feature.
We do not treat the case $G_\phi\subset G_q$, with discrete $G_q$, as we were 
unable to find examples.

Our example model is a $\ZZ_8$ symmetric 4HDM. We have seen in 
Sec.\,\ref{sec:discard} that a $\ZZ_8$ symmetry in the Yukawa sector cannot be 
compatible to a $\ZZ_8$ symmetry in the Higgs potential.
However, we still obtain a fine $\ZZ_8$ symmetric model 
(with regard to the absence of Goldstone bosons)
if the Higgs potential exhibits $\ZZ_8$ symmetry but the Yukawa sector has 
an \textit{accidental} continuous symmetry containing $\ZZ_8$ as a subgroup.

We begin with the scalar sector symmetric by the desired discrete symmetry and 
extend such a symmetry to the other sector. 
A $\ZZ_8$ symmetric Higgs potential is ensured if the phase sensitive (quartic) 
terms in the potential comes from the $D$-matrix
\eq{
\label{Z8-4hdmd:V}
D^V=
\left(
\begin{array}{ccccccc}
 2 & -1 & -1 & 0  \\
 0 & 2 & -1 & -1 \\
 0 & 0 & 2 & -2 \\
\end{array}
\right)
.
}
This matrix has three rows which is the minimum number to sustain a discrete 
symmetry for four fields $\phi_i$, with one automatic $U(1)$ symmetry.
In this case, there are no other phase-sensitive terms except their hermitean 
conjugates.

The next step is to construct the reduced matrix $D^Y$ for the Yukawa 
sector which is compatible with $D^V$ in \eqref{Z8-4hdmd:V}. We need 5 linearly 
independent rows to sustain a discrete symmetry for 7 fields ($\phi_i,Q_{iL}$), 
with two automatic $U(1)$ symmetries.
Therefore, four linearly independent rows are sufficient to sustain an accidental 
$U(1)$ and no more.
Given that each row of $D^Y$ should have a well defined form, we can construct
from the first two rows of $D^V$,
\eq{
\label{U1:ext}
D^Y=
\left(
\begin{array}{cccc|ccc}
 1 & -1 & 0 & 0 & 1 & -1 & 0 \\
 1 & 0 & -1 & 0 & -1 & 1 & 0 \\
 0 & 1 & -1 & 0 & 0 & 1 & -1 \\
 0 & 1 & 0 & -1 & 0 & -1 & 1 \\
\end{array}
\right)\,.
}
We can see that by summing the second row to the first and the fourth row to the 
third we effectively obtain $D^V$ in the first and third rows and there is no other 
way we can eliminate the entries in the $Q_L$-sector.
As other choices -- picking two other rows of $D^V$ or relabeling $Q_{iL}$ 
fields -- are possible, we can see that this procedure does not lead to unique 
extensions.
Nevertheless, the matrix $D^Y_\full$, describing the actual Yukawa terms, can be 
obtained from the procedure outlined in Sec.\,\ref{sec:fullY}.

When we consider $N$ Higgs doublets, with $N>4$, we need $D^Y$ to have $N$ linearly 
independent rows to allow only one additional $U(1)$ symmetry.
The first four rows of $D^Y$ can always be chosen as in \eqref{U1:ext} by using two 
different $\bq$-vectors as, for example, $\bq_1=(1,-1,0)$ and $\bq_3=(0,1,-1)$ in 
\eqref{U1:ext}.
However, additional linearly independent rows cannot be added at will because any 
additional $\bq_i$, $i\ge 5$, will be linearly dependent to $\bq_1$ and $\bq_3$.
If more compatible linearly independent rows can not be added, then more additional 
$U(1)$ symmetries will be present in the Yukawa sector. This is usually the case 
for large symmetries in $D^V$ and $N>4$.
If additional compatible rows can be added in sufficient number, then only one 
accidental $U(1)$ may be present.
The checking may be performed by using the discrete charges of the Higgs doublets.

\subsection{From reduced $D^Y$ to full $D^Y$}
\label{sec:fullY}

We present here a method to construct the full $D^Y_\full$, 
involving all fields of the Yukawa terms ($\phi_a,Q_{iL},d_{iR},u_{iR}$), in 
terms of the reduced $D^Y$, involving only $(\phi_a,Q_{iL})$ (also refereed to as 
a backbone structure).
This is an alternative method to the one presented in Ref.\,\cite{abelian:yuk}, 
where charges where used to reconstruct compatible Yukawa terms.
The method is guaranteed to work for $N=2,3,4,5$ but it is more interesting 
for $N>2$ so that we focus on those cases.
The extended $D$-matrix is not unique in general.

We illustrate the method by constructing the Yukawa interactions of the $\ZZ_4$ 
symmetric 3HDM from the backbone structure \eqref{backbone:z4-3h}.
From the potential part, $D^V$, we extract the two phase sensitive terms in the 
potential and complete with hermitean terms.
From the Yukawa part, $D^Y$, we can construct Yukawa terms.
We first take $D^Y$, with the minimal number of rows
\eq{
\label{backbone:z4-3h:DY}
D^Y=
\left(
\begin{array}{ccc|ccc}
 1 & -1 & 0 & 1 & -1 & 0 \\
 1 & -1 & 0 & -1 & 1 & 0 \\
 0 & 1 & -1 & 0 & 1 & -1 \\
 1 & 0 & -1 & 0 & -1 & 1 \\
\end{array}
\right)\,.
}
For $N=3$, we have $n=N+1=4$ rows in the minimal $D^Y$.
We should associate to each right-handed quark $d_{iR}$ and $u_{iR}$ a row of $D^Y$ 
so that all rows are exhausted. 
For concreteness, we associate 
\eq{
  \label{link}
D^Y=\mtrx{\bd^{(1)}\cr\bd^{(2)}\cr\bd^{(3)}\cr\bd^{(4)}}
\begin{matrix}
    \to& d_{1R}, &u_{2R}\,\, 
  \cr\to& d_{2R}, &u_{3R}\,\,
  \cr\to& d_{3R}&\cr\to&& u_{1R}\,.
\end{matrix}
}
For $N=3,4$, some rows should be simultaneously associated to more than one 
right-handed field. For $N=5$, exactly one right-handed field is paired up with 
exactly one row. For $N\ge 6$, one right-handed field should be linked to more than 
one row and more constraints emerge.

Next, we follow each link in \eqref{link} and build a pair of $D$-matrix 
vectors (rows).
For example, for $\bd^{(1)}\to d_{1R}$ we construct two pairs of vectors,
\eq{
  \label{recipe:d}
\bd^{(1)}\to 
\left\{
\begin{matrix}
\bu^{(1)}(d_{1R})=(e_1;-e_2;e_1;0_3)\cr
\bu^{(1')}(d_{1R})=(e_2;-e_1;e_1;0_3)\cr
\end{matrix}
\right.,
}
where we use the compact notation of canonical vectors $(e_i)_{j}=\delta_{ij}$ and 
follow the ordering $(\phi_a;Q_{iL};d_{iR};u_{iR})$. 
The first vector, for example, reconstructs the Yukawa term 
$\bar{Q}_{2L}\phi_1d_{1R}$, while the second reconstructs 
$\bar{Q}_{1L}\phi_2d_{1R}$. These two terms are the only ones involving $d_{1R}$ 
that satisfies $\bu^{(1)}(d_1)-\bu^{(1')}(d_1)=\bd^{(1)}$.
Since the Yukawa terms for up-type quarks are different, the vectors for them are 
different.
For example, for $\bd^{(1)}\to u_{2R}$ we construct two pairs of vectors,
\eq{
  \label{recipe:u}
\bd^{(1)}\to 
\left\{
\begin{matrix}
\bu^{(1)}(u_{2R})=(-e_1;-e_1;0_3;e_2)\cr
\bu^{(1')}(u_{2R})=(-e_2;-e_2;0_3;e_2)
\end{matrix}
\right.,
}
obeying $\bu^{(1)}(u_{2R})-\bu^{(1')}(u_{2R})=-\bd^{(1)}$. The two vectors 
correspond unambiguously to the Yukawa terms $\bar{Q}_{2L}\tphi_2u_{2R}$ and 
$\bar{Q}_{1L}\tphi_1u_{2R}$.

By applying the recipe to all associations in \eqref{link}, following the order 
$(d_{1R},d_{2R},d_{3R},u_{1R},u_{2R},u_{3R})$, we obtain the $12\times 
n_F $ matrix
\eq{
\label{DY:full:z4-3h}
D^{Y}_{\rm full}=
\left(
\begin{array}{ccc|ccc|ccc|ccc}
 1 & 0 & 0 & 0 & -1 & 0 & 1 & 0 & 0 & 0 & 0 & 0 \\
 0 & 1 & 0 & -1 & 0 & 0 & 1 & 0 & 0 & 0 & 0 & 0 \\
 1 & 0 & 0 & -1 & 0 & 0 & 0 & 1 & 0 & 0 & 0 & 0 \\
 0 & 1 & 0 & 0 & -1 & 0 & 0 & 1 & 0 & 0 & 0 & 0 \\
 0 & 1 & 0 & 0 & 0 & -1 & 0 & 0 & 1 & 0 & 0 & 0 \\
 0 & 0 & 1 & 0 & -1 & 0 & 0 & 0 & 1 & 0 & 0 & 0 \\
\hline
 -1 & 0 & 0 & 0 & 0 & -1 & 0 & 0 & 0 & 1 & 0 & 0 \\
 0 & 0 & -1 & 0 & -1 & 0 & 0 & 0 & 0 & 1 & 0 & 0 \\
 -1 & 0 & 0 & -1 & 0 & 0 & 0 & 0 & 0 & 0 & 1 & 0 \\
 0 & -1 & 0 & 0 & -1 & 0 & 0 & 0 & 0 & 0 & 1 & 0 \\
 -1 & 0 & 0 & 0 & -1 & 0 & 0 & 0 & 0 & 0 & 0 & 1 \\
 0 & -1 & 0 & -1 & 0 & 0 & 0 & 0 & 0 & 0 & 0 & 1 \\
\end{array}
\right)\,.
}
The vectors constructed in \eqref{recipe:d} and \eqref{recipe:u} correspond to 
the rows $1,2,9,10$, respectively.
One can check this matrix still exhibits $\ZZ_4$ symmetry.

We should note that care must be taken when extracting the symmetries of subsectors.
If we take the first (last) 6 rows of \eqref{DY:full:z4-3h} associated to the 
$d_R$-sector ($u_R$-sector) and compute its SNF, we conclude that it exhibits 
a $\ZZ_2\times U(1)$ [$\ZZ_2\times U(1)$] symmetry. This could indicate that 
the d- and u-sectors possess each an accidentally larger symmetry which intersects 
into the $\ZZ_4$ symmetry when both sectors are considered. 
That is not the case as we can compute the $\ZZ_4$ charges in the entire theory and 
list all the compatible Yukawa terms. We discover that in each sector there is one 
term missing which reduces the apparent larger symmetry to $\ZZ_4$ in each sector; 
see more details in appendix \ref{ap:accidentalU1}.
Therefore, it is very important to \textit{consider all compatible terms (rows) 
when analyzing the symmetries of subsectors}.
In general, subsectors may or may not possess larger accidental symmetries.
For example, in appendix \ref{ap:accidentalU1}, we show another $\ZZ_4$-3HDM, 
constructed from the same backbone structure \eqref{backbone:z4-3h}, possessing an 
accidental $U(1)$ symmetry in the $u_R$-sector.
These examples also illustrate the fact that the construction procedure 
\eqref{link} does not lead to a unique $D^Y_\full$.

A remark considering the proposition of Sec.\,\ref{sec:notation} is in order.
The matrix $D^Y_\full$ in \eqref{DY:full:z4-3h} may present an apparent 
contradiction to the proposition as one should be able to write the additional 
d-type row as a linear combination of the 12 rows of $D^Y_\full$ but, as the SNF 
factors change, it cannot be written as a linear combination of the first 6 
rows (d-type) alone.
Both applications of the proposition are correct and the point is that both d- and 
u-type rows of $D^Y_\full$ need to be combined to write the additional d-type row. 
Therefore, the recipe in Sec.\,\ref{min->max} is still valid but one must be 
careful when using fewer rows (restricting to subsectors).

A similar procedure can be applied to backbone structures for $N=4,5$.
Easily identifiable variants can be constructed by relabeling equal-type 
fields. Genuinely different models (different symmetries) can be constructed by 
choosing different rows of the reduced $D^Y$; see example in appendix 
\ref{ap:accidentalU1}.

\section{Discussion and Conclusion}
\label{sec:conclusion}

In this work we have presented techniques to analyze the abelian symmetries in full 
models taking the class of general N-Higgs-doublet models as an example. The 
techniques are based on the Smith normal form and extends the results of 
our previous work\,\cite{abelian:yuk} by focusing on methods to analyze the 
compatibility between abelian symmetries acting in two different sectors -- the 
Yukawa sector and the Higgs potential in our case. Application to other full models 
containing two or more sectors follows analogously.
We have also presented techniques to construct symmetric models by extending the 
symmetry from one sector to another.

The main result within N-Higgs-doublet models is an updated list of abelian 
symmetries that acts compatibly (realizable) in both Yukawa interactions and Higgs 
potential. We focus in faithfully compatible symmetries where the same symmetry 
acts faithfully in both sectors and no larger symmetry is accidentally present in 
either sector. As a result, we have concluded that realizable abelian symmetries 
$G_F$ in the full NHDM should obey the bound
\eq{
  \label{maximal:GF}
|G_F| < \left\{\begin{array}{ll}
    n^2/3\,, & \text{ if $n$ is divisible by $3$,}\cr
    (n^2-1)/3\,, & \text{ if $n$ is not divisible by $3$,}
    \end{array}\right.
}
where $n=N+1$.
Table \ref{table:nhdm} summarizes our results for small $N$.
\begin{table}[htb]
\centering\normalsize
$
\begin{array}{r|ccccc}
N~~ & 2 & 3 & 4 & 5 & 6 \\\hline
|G_q| \le  & 3 & 5 & 8 & 12 & 16 \\
|G_\phi| \le & 2 & 4 & 8 & 16 & 32 \\
|G_F| \le & 2 & 4 & \underline{8} & \underline{12} & \underline{16} \\
\end{array}
$
\caption{\label{table:nhdm}
  Bound for the order of abelian symmetry $G_F=G_q=G_\phi$ acting compatibly on 
full $N$-HDM. $G_q$ and $G_\phi$ denote the symmetries in the Yukawa and Higgs 
potential, respectively.}
\end{table}
The underlined numbers mean that the order of the group should be strictly smaller.
In special, $\ZZ_2$-2HDM is long known to be constructible whereas the 
$\ZZ_2\times\ZZ_2$-3HDM and $\ZZ_4$-3HDM were explicitly shown to 
be realizable in Sec.\,\ref{sec:max:3h}.
We should note that $\ZZ_2\times\ZZ_2$-3HDM models presented here are different 
from the Weinberg model\,\cite{weinberg:scpv} in the Yukawa sector as the charges 
are not flavor universal.

If we allow for accidental symmetries, then the bound \eqref{maximal:GF} does 
not need to be respected. In special, $\ZZ_8$-4HDM is possible: the 
theory possesses a $\ZZ_8$ symmetry in the Higgs potential but an accidental $U(1)$ 
symmetry in the Yukawa sector. Usually, more Higgs doublets and large abelian 
symmetries in the Higgs sector leads to more accidental $U(1)$ symmetries in the 
Yukawa sector.

Based on the results presented here, we can also draw several conclusions 
about the supersymmetric case, i.e., the supersymmetric version of NHDMs with $N$ 
\textit{pairs} of Higgs doublets $H_{ui},H_{di}$, with $i=1,\ldots,N$; we denote 
them as $2N$-HMSSM.
To use our results, we rename $\phi_i=H_{di}$ and 
$\phi_{N+i}=\epsilon H^*_{ui}$. The immediate restrictions are as follows:
\begin{itemize}
\item The Higgs potential cannot realize any discrete rephasing 
symmetry $\ZZ_n$, i.e., either the symmetry is softly broken or it is realized as a 
continuous accidental $U(1)$ symmetry.
This follows because the only phase sensitive terms in the potential come from 
soft-breaking quadratic terms $\phi_i^\dag\phi_j$ which contribute as height 2 
vectors in the $D$-matrix and this kind of matrix does not lead to any 
discrete symmetry (see lemma in Ref.\,\cite{abelian:yuk}).

\item The bound $\boundY$ in Eq.\,\eqref{bounds} now applies separately to the 
down-type ($\bar{d}y^d_iQH_{di}$) and up-type ($\bar{u}y^u_iQH_{ui}$) Yukawa 
interactions (superpotential) since each sector involves separate Higgs (super) 
fields. Within each sector, there are two automatic symmetries corresponding to 
hypercharge and baryon number. In the whole Yukawa interactions, a combination 
of previous symmetries leads to an additional automatic $U(1)$ corresponding to 
Peccei-Quinn symmetry.
There are two reduced matrices \eqref{tD:Y,V} now, one for 
each sector, and they can be made compatible if the charges for $Q$ are compatible.
So the MSSM cannot sustain any discrete symmetry while $\ZZ_3$ is the maximal 
abelian symmetry that is realizable in the Yukawa sector of the extension of MSSM 
with four Higgs doublets (4-HMSSM), provided that $\ZZ_3$ is realizable within 
down- or up-type sector.

\item The previous bound on the symmetry of Yukawa interactions can be evaded 
if we allow for an additional accidental $U(1)$ symmetry in \textit{both} down- 
and up-sectors, so that their intersection leads to a discrete symmetry.
For example, $\ZZ_4$ and $\ZZ_2\times \ZZ_2$ are realizable in 4-HMSSM and they are 
the maximal groups.
For general $2N$-HMSSM, we can show that the symmetry group consists of at 
most two factors $G_q=\ZZ_{k_1}\times\ZZ_{k_2}$ ($k_1$ divides $k_2$) and an 
example with $G_q=\ZZ_N\times\ZZ_N$ or $\ZZ_{N^2}$ can be readily constructed. 
Therefore a group of order $N^2$, which is always larger than the bound $\boundY$ 
in Eq.\,\eqref{bounds}, is always realizable but its maximality remains to be 
checked.

\end{itemize}

In summary, the techniques presented here were shown to be powerful to 
analyze the possible abelian symmetries of full N-Higgs-doublet models 
(and its supersymmetric extensions).
Its capabilities, however, are not restricted to these classes of models and 
further application can be considered on other contexts where discrete or 
continuous symmetries are crucial.
Recent examples can be found in the active areas of neutrino flavor, dark matter 
and axion model building.

\acknowledgments

The author thanks Igor Ivanov and Rabindra Mohapatra for very helpful comments.
This work was partially supported by CNPq, Conselho Nacional de Desenvolvimento 
Científico e Tecnológico - Brasil, and by Brazilian Fapesp through grants 
2013/26371-5 and 2013/22079-8. 

\appendix
\section{Proof of proposition}
\label{ap:teo}

Let us prove the main proposition of Sec.\,\ref{sec:notation}.
Suppose $D$ has size $n\times m$ and rank $n$.
We denote by $D'$ the matrix $D$ after addition of the row 
$\bu=(u_1,u_2,\cdots,u_m)$:
\eq{
D'=\mtrx{    D \\ \hline \bu}\,.
}
The existence of SNF for $D$, $RDC=\snf(D)$, implies
\eq{\label{teo:n+1}
\left(\begin{array}{c|c}
    R &\\ \hline &1
    \end{array}\right)
D'C=
\left(\begin{array}{c|c}
    \diag(d_1,\cdots,d_n) & 0_{m-n} \\ \hline u_1'\hs{1.5em} \cdots \hs{1.5em}u_{n}'
    & \cdots u'_m
    \end{array}\right)\,,
}
where $u'_i=(\bu C)_i$.
Firstly, $u'_1$ should be divisible by $d_1$ because otherwise we could write
\eq{
u'_1=k_1d_1+r_1\,,~~0<r_1<d_1\,,
}
and then we could replace $u'_1$ by its remainder $r_1$ after subtracting a 
multiple of the first row from the last row.
If $r_1$ divides $d_1$, then $r_1$ would be the first factor of the SNF of $D'$ 
which is a contradiction.
If $r_1$ does not divide $d_1$, then a even smaller number could be produced by 
subtraction and consequently leading to a factor in $\snf(D')$ smaller than $d_1$ 
which is a contradiction as well.
From similar arguments, all $u'_i$, $i>1$, should be divisible by $d_1$.
We are then left with 
\eq{
D'\sim\left(\begin{array}{c|c}
    d_1 & 0 \cdots 0\\ \hline 
    \begin{array}{c} 0\\[-1ex]\vdots \\[-.3ex]0\end{array}  &  B
    \end{array}\right)\,,
}
where $B$ has entries divisible by $d_1$ in the same structure as \eqref{teo:n+1}, 
without the first row and column.
Now, the same arguments as above apply to $B$ and the second factor $d_2$.
We can continue until the $d_n$ factor and then all $u'_i$, $1\le i\le n$, can be 
eliminated by row operations. The remaining $u_i'$, $i>n$, should be zero because 
otherwise there would be more nonzero factors in $\snf(D')$.

We conclude that $\bu'=\bu C$ is a linear combination of the rows of 
\eqref{teo:n+1} which implies $\bu$ is a linear combination of the rows of $D$.

Finally, all the arguments above remain valid if the rank of $D$ is smaller than 
its number of rows $n$.

\section{Rephasing space}
\label{ap:remark}

Here we clarify the distinction between the groups $\{\tG_q,\tG_\phi\}$ and
$\{G_q,G_\phi\}$, and stress that the former enter in $G_F=\tG_q\cap\tG_\phi$.
The notation $\tG$ refers to the symmetry group considered as a subgroup 
of $[U(1)]^{n_F}$, the rephasing symmetry of all $n_F$ fields involved.
We present simple examples below where the distinction is important.
In the text, we have disregarded automatic symmetries in $G_q$ and $G_\phi$.

Let us adopt the following notation: we consider a theory consisting of two sectors 
described by Lagrangians $\lag_{1,2}$, with symmetries denoted either as $G_{1,2}$ 
when restricted to lagrangians $\lag_{1,2}$ or as $\tG_{1,2}$ when considering all 
fields appearing in the whole theory ($\lag_1+\lag_2$).

For the first example, consider two noninteracting complex scalar fields 
$\varphi_{1,2}$ and a theory where the only phase sensitive terms are
\eq{
\lag_1=\varphi_1^4\,,~~~~
\lag_2=\varphi_2^3.
}
We have suppressed coupling coefficients and hermitean conjugates for simplicity.
It is clear that $\lag_1$ ($\lag_2$) sustains a $\ZZ_4$ ($\ZZ_3$) symmetry when 
restricted to $\varphi_1$ ($\varphi_2$) sector so that $G_1=\ZZ_4$ ($G_2=\ZZ_3$).
However, to analyze the compatibility of symmetries (Sec.\,\ref{sec:match}) and 
extract the common symmetry of the whole theory, $G_F$, we need to consider
$\tG_1=\ZZ_4\times U(1)_{2}$ and $\tG_2=U(1)_{1}\times\ZZ_3$ which are subgroups of 
the whole rephasing group $[U(1)]^2=U(1)_1\times U(1)_2$ for two fields.
Clearly $G_F=\tG_1\cap\tG_2=\ZZ_4\times\ZZ_3$, although $G_1\cap G_2=\{e\}$ 
abstractly.
This situation can be easily detected in the $D$-matrix as it would be 
separable into blocks.

For the second example, we consider a theory with the presence of an automatic 
\textit{discrete} symmetry.
We consider two chiral fermion fields $\psi_{1,2}$ and one complex scalar field 
$\varphi$.
The two sectors consist of phase sensitive terms
\eqali{
\label{lag:psi}
  -\lag_1&=\overline{\psi_1^c}\psi_1+\overline{\psi_1^c}\psi_2\varphi
    +\overline{\psi_2^c}\psi_2\varphi^*\,,
\cr
  \lag_2&=\varphi^3\,.
}
We can see this theory is invariant by $G_F=\ZZ_6$ symmetry:
\eq{
\label{example:Z6}
\psi_1\to -\psi_1,~~
\psi_2\to e^{i2\pi/6}\psi_2,~~
\varphi\to e^{i2\pi/3}\varphi\,.
}
However, the scalar sector is only sensitive to the $\ZZ_3$ subgroup of $\ZZ_6$.
The $\ZZ_2$ subgroup is automatic and it is a discrete version of fermion number: 
$\psi_{1,2}$ are odd whereas $\varphi$ is even. 
The symmetry $\ZZ_6$ is the intersection of $\tG_2=\ZZ_3\times[U(1)]^2$ and 
$\tG_1=G_1=\ZZ_6$, although clearly $G_2=\ZZ_3$.
In the SNF method, the information on the automatic $\ZZ_2$ can be seen on the 
$\ZZ_6$-charge,
\eq{
\bss=(3,1,2)\,,
}
with ordering $(\psi_1,\psi_2,\varphi)$, as the scalar charge has a factor 
that divides $|G_F|=6$.
The same happens if the charges for a set of fields (e.g. scalars) has a common 
factor $k$ which divides the whole group.
See more examples of models with automatic discrete lepton number in 
Ref.\,\cite{serodio:nu.texture}. In particular, a type-II seesaw model presents 
$G_1=\ZZ_3$ and $G_2=\ZZ_3$ but $G_F=\ZZ_6$.

\section{Backbone structure for maximal symmetry in 3HDM}
\label{ap:order4-3hdm}

Let us show that the backbone structures shown in \eqref{backbone:z4-3h} and
\eqref{backbone:z2z2-3h}, corresponding to the order 4 abelian symmetries $\ZZ_4$
and $\ZZ_2\times\ZZ_2$ respectively, are unique up to reordering of $\phi_a$ and
$Q_{iL}$ fields.

The minimal $D^Y$ matrix for 3HDM has the form in \eqref{DY:3h}.
In this case, the $\bp$- and $\bq$-vectors are any of the six vectors that can be
categorized into the three types of nonparallel vectors in Eq.\,\eqref{q-vectors}.
We have seen in \cite{abelian:yuk} that $|G_q|$ is bounded above by
$m=n_1n_2+n_2n_3+n_3n_1$ where $n_i$ is the number of $i$-th type $\bq$-vector (or,
in this case, $\bp$-vector as well).
To get $|G_q|=4$, we can allow $(n_1n_2n_3)=(211)$ or $(n_1n_2n_3)=(220)$.
Therefore, we have at least one and at most two $\bq$-vectors ($\bp$-vectors) 
parallel (of the same type).

We know that the order of the abelian group, $|G_F|$, can be
given by the modulus of a determinant-like $E$-function which is a function of the
rows of the $D$-matrix\,\cite{abelian:yuk}. Such
function can be expanded as
\eq{
\label{E:3h:order4}
E(D^Y)=E(\bp_1+\bq_1,\cdots,\bp_4+\bq_4)
=(\slashed{1}\slashed{2}34)-(\slashed{1}\slashed{2}31)+(\slashed{1}\slashed{3}21)
 -(\slashed{1}\slashed{3}24)-(\slashed{3}\slashed{2}14)
\,,
}
where we use the shorthand: $\bp_i\to i$, $\bq_k\to\slashed{k}$ and suppress
the $E$ function so that the first term represent
$E(\bq_1,\bq_2,\bp_3,\bp_4)$. We also use determinant-like properties for $E$; see
Ref.\,\cite{abelian:yuk} for more details.
We also conventionally choose $\bq_4=\bq_1$ to be the parallel vectors.
And then $\bq_2$ and $\bq_3$ cannot be parallel to $\bq_1$.
We are left with two possibilities: either $\bq_3$ is parallel to $\bq_2$ or not.
To avoid cancellations in \eqref{E:3h:order4}, we need to impose $\bp_4\neq\bp_1$
as well.
From the properties of $E$, one can also check that exchanging $\bp_1\leftrightarrow
\bp_4$ only amounts to an overall sign change.

Recall that any of the five terms in \eqref{E:3h:order4} is $\pm 1$ or $0$, the
latter being possible only if some of the two $\bp$-vectors or $\bq$-vectors are
parallel.
To get $|E(D^Y)|=4$, we need exactly one vanishing term.
Each of the first four terms only vanish if its $\bp$-vectors are parallel.
The last term in \eqref{E:3h:order4} vanishes if one (or both) of the pairs
$\{\bq_2,\bq_3\}$ or $\{\bp_1,\bp_4\}$ is parallel. 
In the first case, we can conventionally adopt $\bq_3=\bq_2$ and, among the
$\bp$-vectors, only $\{\bp_1,\bp_4\}$ or $\{\bp_2,\bp_3\}$ may be parallel to avoid
the first four terms from vanishing. However, since $\bq_4=\bq_1$ and
$\bq_3=\bq_2$, parallel $\{\bp_2,\bp_3\}$ is equivalent to parallel
$\{\bp_1,\bp_4\}$ after we exchange $\bp_2\leftrightarrow\bp_1$ and
$\bp_3\leftrightarrow\bp_4$.

If $\bq_3,\bq_2$ are not parallel, then we must choose some other pair of
$\bp$-vectors to be parallel.
We have three possibilities: $\bp_2=\pm\bp_1,\bp_3=\pm\bp_1,\bp_4=\pm\bp_1$.
The possibility that $\bp_2$ or $\bp_3$ is parallel to $\bp_4$ is taken into
account from the exchange symmetry $\bp_1\leftrightarrow\bp_4$.
The possibility $\bp_3=\pm\bp_1$ can be also taken into account by exchanging
the second and third row of $D^Y$, i.e., $\bq_2\leftrightarrow\bq_3$
and $\bp_2\leftrightarrow\bp_3$.

Summarizing all possibilities, we are left with the following cases
\eqali{
\text{(A)~~}&\bq_3=\bq_2 \text{ and } \bp_4=\pm\bp_1;\cr
\text{(B)~~}&\{\bq_2,\bq_3\} \text{ nonparallel and } \bp_2=\pm\bp_1;\cr
\text{(C)~~}&\{\bq_2,\bq_3\}  \text{ nonparallel and } \bp_4=\pm\bp_1.
}
We analyze them in detail in the following.

For case (A), the expansion of \eqref{E:3h:order4} becomes
\eqali{
\label{E:3h:order4:A}
E(D^Y)&=\pm(\slashed{1}\slashed{2}31)-(\slashed{1}\slashed{2}31)
  +(\slashed{1}\slashed{2} 21) \mp(\slashed{1}\slashed{2}21)
\,,\cr
&=-2(\slashed{1}\slashed{2}31) +2(\slashed{1}\slashed{2} 21)\,,
}
where we chose $\bp_4=-\bp_1$ in the second line to avoid cancellations.
Further cancellation is avoided only if $D^Y$ has one of the forms
\eq{
\label{DY:3h:A}
D^Y_{A.1}=\left(\begin{array}{c|c}
  \bp_1 & \bq_1\cr
  \bp_2 & \bq_2\cr
  -\bp_2 & \bq_2\cr
  -\bp_1 & \bq_1
  \end{array}\right)
\text{~~or~~}
D^Y_{A.2}=
\left(\begin{array}{c|c}
  \bp_1 & \bq_1\cr
  \bp_2 & \bq_2\cr
  -\bp_2\pm\bp_1 & \bq_2\cr
  -\bp_1 & \bq_1
  \end{array}\right)\,,
}
where the sign in the last case should be chosen so as to make
$\bp_3=-\bp_2\pm\bp_1$ a $\bp$-vector.

For case (B), we expand $\bq_3=\bq_1+\bq_2$ by choosing the signs appropriately.
The expansion of \eqref{E:3h:order4} reads
\eqali{
\label{E:3h:order4:B}
E(D^Y)&=
(\slashed{1}\slashed{2}34)-(\slashed{1}\slashed{2}31)
 \mp(\slashed{1}\slashed{2}14)-(\slashed{1}\slashed{2}14)
\,,\cr
&=(\slashed{1}\slashed{2}34)+(\slashed{1}\slashed{2}13)
  -2(\slashed{1}\slashed{2}14)\,,
}
where we chose $\bp_2=\bp_1$ in the second line to avoid cancellations.
There is no cancellation if $\bp_3=-\bp_4-\bp_1$ and we get
\eq{
\label{DY:3h:B}
D^Y_B=\left(\begin{array}{c|c}
  \bp_1 & \bq_1\cr
  \bp_1 & \bq_2\cr
  -\bp_1-\bp_4 & \bq_1+\bq_2\cr
  \bp_4 & \bq_1
  \end{array}\right)\,.
}
Note that $\bp_3=-\bp_4-\bp_1$ should be a $\bp$-vector.

For case (C), we expand again $\bq_3=\bq_1+\bq_2$.
The expansion of \eqref{E:3h:order4} reads
\eqali{
\label{E:3h:order4:C}
E(D^Y)&=
 \pm(\slashed{1}\slashed{2}31)-(\slashed{1}\slashed{2}31)+(\slashed{1}\slashed{2}21)
 \mp(\slashed{1}\slashed{2}21)
\,,\cr
&=-2(\slashed{1}\slashed{2}31)  +2(\slashed{1}\slashed{2}21)\,,
}
where we need to choose $\bp_4=-\bp_1$ in the second line.
The necessary form for $\bp_3$ can be extracted and we obtain
\eq{
\label{DY:3h:C}
D^Y_{C.1}=\left(\begin{array}{c|c}
  \bp_1 & \bq_1\cr
  \bp_2 & \bq_2\cr
  -\bp_2 & \bq_1+\bq_2\cr
  -\bp_1 & \bq_1
  \end{array}\right)
\text{~~or~~}
D^Y_{C.2}=\left(\begin{array}{c|c}
  \bp_1 & \bq_1\cr
  \bp_2 & \bq_2\cr
  -\bp_2\pm\bp_1 & \bq_1+\bq_2\cr
  -\bp_1 & \bq_1
  \end{array}\right)\,,
}
where all rows should correspond to allowed $\bp$- and $\bq$-vectors.

By choosing $\bp_1=\bq_1=(1,-1,0)$ and $\bp_2=\bq_2=(0,1,-1)$ we can rewrite
\eqref{DY:3h:A} and \eqref{DY:3h:C} as
\eq{
\label{DY:3h:Z4}
D^Y_{A.2}=
\left(
\begin{array}{ccc|ccc}
 1 & -1 & 0 & 1 & -1 & 0 \\
 0 & 1 & -1 & 0 & 1 & -1 \\
 -1 & 0 & 1 & 0 & 1 & -1 \\
 -1 & 1 & 0 & 1 & -1 & 0 
\end{array}
\right),
~~
D^Y_{C.1}
=
\left(
\begin{array}{cccccc}
 1 & -1 & 0 & 1 & -1 & 0 \\
 0 & 1 & -1 & 0 & 1 & -1 \\
 0 & -1 & 1 & 1 & 0 & -1 \\
 -1 & 1 & 0 & 1 & -1 & 0 
\end{array}
\right),
}
and
\eq{
\label{DY:3h:Z2Z2}
D^Y_{A.1}=
\left(
\begin{array}{cccccc}
 1 & -1 & 0 & 1 & -1 & 0 \\
 0 & 1 & -1 & 0 & 1 & -1 \\
 0 & -1 & 1 & 0 & 1 & -1 \\
 -1 & 1 & 0 & 1 & -1 & 0 
\end{array}
\right),~~
D^Y_{C.2}=
\left(
\begin{array}{cccccc}
 1 & -1 & 0 & 1 & -1 & 0 \\
 0 & 1 & -1 & 0 & 1 & -1 \\
 -1 & 0 & 1 & 1 & 0 & -1 \\
 -1 & 1 & 0 & 1 & -1 & 0
\end{array}
\right).
}
We can see they all have three rows in common.
In fact the matrices in \eqref{DY:3h:Z4} are compatible with the $\ZZ_4$ charges
in \eqref{3h-z4:charge}, i.e., upon multiplication on the charges, it gives 4 times
a vector of relatively prime integers.
Analogously, the matrices in \eqref{DY:3h:Z4} are compatible with the
$\ZZ_2\times\ZZ_2$ charges in \eqref{3h-z2z2:charge}.

For case (B), by choosing $\bp_1=\bq_1=(1,-1,0)$ and $\bq_2=\bp_4=(0,1,-1)$, 
Eq.\,\eqref{E:3h:order4:B} reads
\eq{
D^Y_B=
\left(
\begin{array}{cccccc}
 1 & -1 & 0 & 1 & -1 & 0 \\
 1 & -1 & 0 & 0 & 1 & -1 \\
 -1 & 0 & 1 & 1 & 0 & -1 \\
 0 & 1 & -1 & 1 & -1 & 0
\end{array}
\right)\,.
}
The SNF form of this matrix reveals a $\ZZ_4$ symmetry which is not compatible
with the matrices in \eqref{DY:3h:Z4}.
But if we choose $\bp_1=\bq_1=(1,0,-1)$ and $\bq_2=\bp_4=(0,-1,1)$, we obtain
\eq{
\label{DY:3h:Z4:B}
D^Y_{B}=
\left(
\begin{array}{cccccc}
 1 & 0 & -1 & 1 & 0 & -1 \\
 1 & 0 & -1 & 0 & -1 & 1 \\
 -1 & 1 & 0 & 1 & -1 & 0 \\
 0 & -1 & 1 & 1 & 0 & -1 
\end{array}
\right)\,,
}
which is compatible with \eqref{DY:3h:Z4} and the charges \eqref{3h-z4:charge}.

Therefore, \textit{there is only one backbone structure} for 3HDM with symmetries
$\ZZ_4$ or $\ZZ_2\times\ZZ_2$.
To obtain the backbone structure \eqref{backbone:z4-3h}, we just collect
the nonequivalent rows of the matrices in \eqref{DY:3h:Z4} and \eqref{DY:3h:Z4:B}.
Analogously, the backbone structure \eqref{backbone:z2z2-3h} is extracted from 
\eqref{DY:3h:Z2Z2}.

\section{Height of $\bp'_i$ vectors for maximal $D^Y$}
\label{ap:pi'}

We start from the requirement for maximal $G_F$:
\eq{
E(\bp'_3,\bp'_4,\cdots,\bp'_n)=\pm\boundY\,.
}
The determinant-like $E$-function was defined in Ref.\,\cite{abelian:yuk} and 
coincides with the determinant of $D^Y$ after eliminating the first and last 
columns. When $D^Y$ is in the form \eqref{tD:Y:2}, we conclude that only the 
rows corresponding to $\bp'_i$ contributes non-unit factors.
The properties of the $E$-function allows us to expand it using Eqs.\,\eqref{pi'} 
and \eqref{p3'} as
\eqali{
  \label{E-expansion}
E(\bp'_3,\bp'_4,\cdots,\bp'_n)&=E(\bp_3,\bp_4,\cdots,\bp_n)+
    E(-\bp_1,\bp_4,\cdots,\bp_n)+E(-\bp_2,\bp_4,\cdots,\bp_n)
    \cr&~~+\ \text{combinations of }[(\bp_{3k+1}\to -\bp_1) \text{ or } 
    (\bp_{3k+2}\to -\bp_2) \text{ or } (\bp_{3k}\to -\bp_3)]\,.
}
We know that each term in the expansion is either $0$ or $\pm 1$ for height-2 
vectors and so we can write
\eq{
  \label{cond:E<=1}
|E(\{\text{$n$ height-2 vectors}\})|\le 1\,,
}
where height-2 vectors denote only $\bp$-vectors in this case.
To obtain a maximal value, all terms in the expansion should be $\pm 1$ and no 
cancellations should occur. In this case, the maximal value \eqref{bounds} is 
attained.

Using the maximality of the expansion in \eqref{E-expansion}, we can prove that 
$\bp_i'$, $i\ge 4$, are always height-4 vectors.
As an example, we calculate
\eq{
  \label{E:p4'}
E(\bp_3,\bp_4',\bp_5,\cdots,\bp_n)=
E(\bp_3,\bp_4,\bp_5,\cdots,\bp_n)+E(\bp_3,-\bp_1,\bp_5,\cdots,\bp_n),
}
where only $\bp_4'$ is given by \eqref{pi'}.
The two terms in the expansion of \eqref{E:p4'} should add to $\pm 2$ since they 
are also present in the expansion \eqref{E-expansion} which should be maximal.
Condition \eqref{cond:E<=1} tells us that $\bp_4'$ should have height $h(\bp_4')> 
2$ but its expansion $\bp_4'=\bp_4-\bp_1$, a combination of two $\bp$-vectors, 
implies it has height 4.
Analogous arguments show that all $\bp_i'$, $i\ge 4$, are height-4 vectors.
We also arrive at the general rule
\eq{
  \label{cond:E<=2}
|E(\text{height-4 vector},\{\text{$n-1$ height-2 vectors}\})|\le 2\,,
}
where a height-4 vector is restricted to a vector that can be written as a sum of 
two $\bp$-vectors.
Note that the denomination ``sum of two $\bp$-vectors'' has less meaning than its 
height since the sum of two $\bp$-vectors (height-2) can still be a 
$\bp$-vector (height-2) as
\eq{
(1,-1,0,\cdots)+(0,1,-1,0,\cdots)=(1,0,-1,0,\cdots).
}

For $\bp_3'$, we calculate
\eq{
  \label{E:p3'}
E(\bp_3',\bp_4,\bp_5,\cdots,\bp_n)=
  E(\bp_3,\bp_4,\bp_5,\cdots,\bp_n)+E(-\bp_1,-\bp_1,\bp_5,\cdots,\bp_n)
  +E(-\bp_2,\bp_4,\bp_5,\cdots,\bp_n)\,,
}
and confirm they also appear in \eqref{E-expansion}.
Maximality of \eqref{E-expansion}, \eqref{cond:E<=1} and the form of 
$\bp_3'=\bp_3-\bp_1-\bp_2$ implies $\bp_3'$ is a height-6 vector.

\section{Height of $\bp'_3$ can not be reduced}
\label{ap:p3'}

Here we prove that, for maximal $D^Y$, row operations on $\{\bp_i'\}$ in 
Eq.\,\eqref{tD:Y:2} cannot transform the whole set into a generic set $\{\br_i\}$ 
in 
Eq.\,\eqref{tD:Y,V}, which is composed of rows of at most height 4.
We make repeated use of 

\medskip
\begin{minipage}{.9\textwidth}\smallskip
\noindent\textbf{Proposition:} For vectors $\bp_i$, of height 2 for which 
$E(\{\bp_i\})=\pm 1$, and $\bu$, an integer linear combination of $\bp_i$, if 
$|E(\bp_1,\cdots,\bu,\cdots,\bp_n)|\ge k$, then $\bu$ has height $h(\bu)\ge 
2k$.
\end{minipage}\medskip
\\

We prove the equivalent statement:
For vectors $\bp_i$, of height 2 for which $E(\{\bp_i\})=\pm 1$, and $\bu$, an 
integer linear combination of $\bp_i$, 
if $\bu$ has height $h(\bu)<2k$, then $|E(\bp_1,\cdots,\bu,\cdots,\bp_n)|< k$.
This statement is the generalization of the rules \eqref{cond:E<=2} and 
\eqref{cond:E<=1}, without the equality option.
The equality can be included because the height of linear combinations of 
$\bp$-vectors only increases in steps of two.

The proof follows by induction. The case $k=2$ is evident in \eqref{cond:E<=1}.
Now we assume the statement is valid for a vector $\bu'$ of height 
$h(\bu')<2(k-1)$. The statement for a vector $\bu=\bu'+\bp$, of height $h(\bu)<2k$,
is true because
\eqali{
|E(\bp_1,\cdots,\bu,\cdots,\bp_n)|&=|E(\bp_1,\cdots,\bu'+\bp,\cdots,\bp_n)|\cr
&\le |E(\bp_1,\cdots,\bu',\cdots,\bp_n)|+|E(\bp_1,\cdots,\bp,\cdots,\bp_n)|\cr
&<(k-1)+1=k\,.
}
The vector $\bp$ is any $\bp$-vector of height 2.

Now, take $\bp_3''=\bp_3'-\bp_4'=\bp_3-\bp_2-\bp_4$ and calculate
\eqali{
E(\bp_3'',-\bp_1,\bp_5,\cdots,\bp_{n})
&=E(\bp_3,-\bp_1,\bp_5,\cdots,\bp_{n})+E(-\bp_2,-\bp_1,\bp_5,\cdots,\bp_{n})
    +E(-\bp_4,-\bp_1,\bp_5,\cdots,\bp_{n})
\cr
&=E(\bp_3,-\bp_1,\bp_5,\cdots,\bp_{n})+E(-\bp_2,-\bp_1,\bp_5,\cdots,\bp_{n})
    +E(-\bp_1,\bp_4,\bp_5,\cdots,\bp_{n})
\cr
&=\pm 3,
}
for maximal $D^Y$. The proposition tell us that $\bp_3''$ has at least height 6.
Changing the sign, $\bp_3''=\bp_3'+\bp_4'$ make things worse and we can conclude 
that $\bp_3''$ have even greater height. Next, we note that all $\bp_{2k+1}$ are 
all interchangeable to $\bp_4$ and the same conclusion follows for 
$\bp_3''=\bp_3'\mp\bp_{3k+1}$.
Generalization to all $\bp_3''=\bp_3'\mp\bp_{3k+2}$ or $\bp_3''=\bp_3'\pm\bp_{3k}$ 
follows analogously. For the latter, we can calculate 
$E(\bp_3'+\bp_6',\bp_4,\bp_5,-\bp_3,\cdots,\bp_n)$.
Since any row operation on the row $\bp_3'$ in \eqref{condition:full} is generated 
by one of the previous replacements $\bp_3'\to \bp_3''$, we arrive at our desired 
conclusion: we can not decrease the height of $\bp_3'$ by using row operations on 
$D^Y$.
Therefore, the equivalence \eqref{condition:full} is not possible for maximal $D^Y$.

\section{Examples of full models and textures}
\label{ap:accidentalU1}

By using the methods of Sec.\,\ref{sec:extension}, we can construct full models, 
including the explicit Yukawa terms which gives rise to \textit{textures} in the 
quark mass matrices.
We show here some examples to illustrate that $D^Y_\full$ is not univocally 
constructed from $D^Y$.
This is a general feature when we try to extend a symmetry from a 
subsector of a theory to a larger sector involving more fields. In 
particular, this caveat applies to the methods of Secs. \ref{sec:V->Y} and 
\ref{sec:fullY}.

The explicit example of Sec.\,\ref{sec:fullY} constructed the matrix $D^Y_\full$, 
shown in \eqref{DY:full:z4-3h}, from the backbone structure in 
Eq.\,\eqref{backbone:z4-3h}.
The procedure was illustrated in Eq.\,\eqref{link}, where the rows $(1,2,3,4,1,2)$ 
of $D^Y$ were associated to the righthanded quarks 
$(d_{1R},d_{2R},d_{3R},u_{1R},u_{2R},u_{3R})$, in this order. 
If we extract the $\ZZ_4$ charge from $D^Y_\full$ we obtain
\eq{
\bss=(-1,1,0;1,-1,0;0,2,-1;-1,0,2)\,.
}
By using the charge vector, we can obtain all the Yukawa interactions which 
can be collectively summarized by
\eq{
   \label{texture:1}
\bar{Q}_{iL}d_{jR}\sim 
  \mtrx{v_2&v_1&0\cr v_1&v_2&v_3\cr v_3&0&v_2}
  ,~
\bar{Q}_{iL}u_{jR}\sim 
  \mtrx{0&v_1&v_2\cr v_3&v_2&v_1\cr v_1&v_3&0}  
  \,.
}
The presence of $v_3=\aver{\phi_3^{(0)}}$ in the (23) entry of the first matrix 
denotes the coupling of $\phi_3$ with $\bar{Q}_{2L}d_{3R}$; for up quark terms, 
$\tilde{\phi}_i$ should be considered. The two matrices also give the 
order of magnitude of respective quark mass matrices.
Note that the terms corresponding to the (31) and (32) entries in the first and 
second matrices, respectively, are not present in \eqref{DY:full:z4-3h}.
These are the missing terms that ensures the $\ZZ_4$ symmetry in both d- and 
u-sectors separately.
We can see that this $\ZZ_4$ symmetry gives rise to two-zero 
textures\,\cite{grimus:texture}.

As a second example, we can construct another $\ZZ_4$-3HDM from the same backbone 
\eqref{backbone:z4-3h}.
In this case, we consider rows $(1,2,3,4,5,6)$ of $D^Y$ and associate to 
right-handed quarks $(d_{1R},d_{2R},d_{3R},u_{1R},u_{2R},u_{3R})$, in the same 
order. 
The matrix $D^Y_\full$ has 12 rows that can be separated in two submatrices of 6 
rows, one for the d-sector and another for the u-sector.
Each submatrix leads us to conclude that the d-sector has apparent $\ZZ_2\times 
U(1)$ symmetry and the u-sector exhibits apparent $[U(1)]^2$ symmetry.
The $\ZZ_4$ charge obtained from $D^Y_\full$,
\eq{
  \label{Z4:case2}
\bss=(1,-1,0;-1,1,0;0,2,1;1,-1,0)\,,
}
can be used to find one additional term for each d- and u-sectors.
The complete Yukawa terms can be summarized in
\eq{
 \label{texture:2}
\bar{Q}_{iL}d_{jR}\sim
  \mtrx{v_2&v_1&0\cr v_1&v_2&v_3\cr v_3&0&v_2}
  ,~
\bar{Q}_{iL}u_{jR}\sim
  \mtrx{0&v_3&v_1\cr v_3&0&v_2\cr v_1&v_2&v_3}\,.
}
The terms that are initially missing correspond to the (31) and (23) entries in the 
first and second matrices, respectively.
With the addition of these terms, we find a $\ZZ_4$ symmetry in the d-sector and an 
accidental $U(1)$ symmetry in the u-sector.
The presence of an accidental symmetry signals that this second model is 
essentially different from the first.
The symmetries $\ZZ_4$ and $U(1)$ can be checked by using the common 
charge \eqref{Z4:case2}.
One can further see that the textures in Eqs.\,\eqref{texture:1} and 
\eqref{texture:2} are still general and can accommodate the experimental 
values for quark masses and mixing parameters since these texture-zeros
can be achieved through weak basis change and they impose no physical 
restriction\,\cite{xing}.


\end{document}